\documentclass[aps,twocolumn,showpacs,superscriptaddress,pra,10pt]{revtex4-1}
\usepackage{graphicx,braket,amsmath,amsfonts,color}
\usepackage[normalem]{ulem}
\usepackage[bookmarks=true,colorlinks=true,breaklinks]{hyperref}

\usepackage{epstopdf}

\newcommand*\ham{\hat{H}}
\newcommand*\af{\gamma}
\newcommand*\afB{\delta}
\newcommand*\afpi{\gamma}
\newcommand*\crt[1]{\hat{a}^\dagger_{#1}}
\newcommand*\dst[1]{\hat{a}^{\phantom{\dagger}}_{#1}}
\newcommand*\vett[1]{{\bf{#1}}}

\newcommand*\oper[1]{ \hat{#1} }
\newcommand{\kpt}{ \vett{k} }

\newcommand{\sym}{ s }
\newcommand{\symgroup}{ S }
\newcommand{\angstrom}{\mbox{\normalfont\AA}}

\begin{document}

\author{Mario Motta}
\affiliation{Division of Chemistry and Chemical Engineering, California Institute of Technology, 
Pasadena, CA 91125, USA}

\author{Shiwei Zhang}
\affiliation{Center for Computational Quantum Physics, Flatiron Institute, New York, NY 10010, USA}
\affiliation{Department of Physics, College of William and Mary, Williamsburg, VA 23187-8795, USA}

\author{Garnet Kin-Lic Chan}
\affiliation{Division of Chemistry and Chemical Engineering, California Institute of Technology,
Pasadena, CA 91125, USA}

\title{Hamiltonian symmetries in auxiliary-field quantum Monte Carlo \\
        calculations for electronic structure}

\begin{abstract}
We describe how to incorporate symmetries of the Hamiltonian into auxiliary-field quantum Monte Carlo
calculations (AFQMC).
Focusing on the case of Abelian symmetries, we show that the computational cost of most steps of an 
AFQMC calculation is reduced by $N_k^{-1}$, where $N_k$ is the number of irreducible representations
of the symmetry group.
We apply the formalism to a molecular system as well as to several crystalline solids.
In the latter case, the lattice translational group provides increasing savings
as the number of $k$ points is increased, which is important in enabling calculations
that approach the thermodynamic limit. 
The extension to non-Abelian symmetries is briefly discussed.
\end{abstract}

\maketitle

\section{Introduction}

The basic task of electronic structure (ES) theory is to solve the time-independent Schr\"odinger 
equation $\oper{H} \ket{\Psi_\mu} = E_\mu \ket{\Psi_\mu}$, to determine the 
eigenvalues $E_\mu$ and eigenstates $\ket{\Psi_\mu}$ of a Hamiltonian $\oper{H}$.
When relativistic effects and nuclear motion are neglected, the second-quantized Born-Oppenheimer 
Hamiltonian operator has the form \cite{Born_1927,Ziman_book_1965,Szabo_book_1989},  
\begin{equation}
\begin{split}
\label{eq:ham2nd}
\ham 
&= \sum_{ pq } h_{pq} \crt{p} \dst{q} 
+ 
\frac{1}{2} 
\sum_{ pqrs }
(pr|qs) \, \crt{p} \crt{q} \dst{s} \dst{r} \,\, , \\
h_{pq} &= \int d{\bf r} \, \varphi^*_p({\bf r})\left(- \frac{1}{2} \nabla^2-
\sum_{a} \frac{Z_{a}}{|{\bf r}-{\bf R}_{a}|} \right)\varphi_q^{\,}({\bf r})  \,\, , \\
(pr|qs) &=
\int d{\bf r} \, d{\bf r}^{\prime} \, \varphi^*_p({\bf r})\varphi^{\,}_r\left({\bf r}\right) \,
\frac{ 1 }{|{\bf r}-{\bf r}^{\prime}|} \, \varphi^*_q({\bf r}^{\prime})\varphi_s^{\,}\left({\bf r}^{\prime}\right) \,\, , \\
\end{split}
\end{equation}
expressed here in atomic units, using 
a basis of $M$ spin-orbitals $\{ \varphi_{p} \}_{p=1}^M$ for the one-electron Hilbert space.
In Eq. \eqref{eq:ham2nd}, $a$ labels nuclei with positions ${\bf R}_{a}$, and ${\bf r}, {\bf r}^\prime$ denote electronic coordinates.

Many ES methods take advantage of Hamiltonian symmetries~\cite{dupuis1977molecular,Szabo_book_1989,Stanton_JCP_1991,chan2002highly,Berkelbach_JCTC_2017,Sun_WIRES_2018} 
to improve the efficiency of calculations.
In the present work, we give an instructional account of how to
incorporate symmetry into the auxiliary-field quantum Monte Carlo (AFQMC) 
method~\cite{Blankenbecler_PRD_1981,Sugiyama_Annals_1986,Zhang_PRB55_1997,Rom_JCP_1998,Baer_JCP_1998,Zhang_PRL90_2003,AlSaidi_JCP124_2006,Zhang_Notes_2013,Motta_WIRES_2018} 
through symmetry adapted orbitals~\cite{dupuis1977molecular,Szabo_book_1989}.
We illustrate the formalism through the use of reflection symmetry in molecules
and lattice translational symmetry in periodic solids. In the case of lattice translational symmetry,
the symmetry orbitals become crystalline orbitals, and thus the AFQMC calculation
is carried out in a similar framework to other recent implementations of quantum chemical many-body methods in crystals~\cite{Evarestov_book,dovesi2014crystal14,booth2016plane,Berkelbach_JCTC_2017,sun2017gaussian,Sun_WIRES_2018,gruber2018applying}.
This use of translational symmetry should be distinguished from twist averaging ~\cite{lin2001twist,Ma_PRL114_2015,Morales_arxiv_2018a,Morales_arxiv_2018b}, 
as incorporating larger translational groups reduces many-body size effects, rather 
than only the one-body size effects captured by twist averaging.

The remainder of the paper is structured as follows. In Section \ref{sec:methods} we briefly review the 
connection between Hamiltonian symmetries, matrix element sparsity, and
the AFQMC formalism. We then describe in detail how to use Hamiltonian symmetries 
to decrease the cost of the various operations involved in an AFQMC calculation.
In Section \ref{sec:results} we apply the formalism to
a test molecule, using
reflection symmetries, and crystalline solids, with
increasing sizes of the translational group. Conclusions are drawn 
in Section \ref{sec:conclusions}. Further implementation details are provided in the Appendices.

\section{Background}
\label{sec:methods}

A transformation operator $\oper{\sym}$ is a symmetry operator for a 
Hamiltonian $\oper{H}$ if the latter is invariant under the transformation $\oper{s}$,
\begin{equation}
[ \oper{\sym} , \oper{H} ] = 0 \quad .
\label{eq:smy_group}
\end{equation}
The set of operators $\oper{\sym}$ such that \eqref{eq:smy_group} holds, forms a group
$\mathcal{\symgroup}$ under composition, termed the symmetry group of $\hat{H}$.
Here, we will focus on Abelian symmetries, i.e. we will assume $[ \oper{\sym}_1 , 
\oper{\sym}_2 ] = 0$ for all $\oper{\sym}_1, \oper{\sym}_2 \in \mathcal{\symgroup}$.

\subsection{Hamiltonian symmetries and sparsity} 
\label{sec:sparse}

In the one-electron Hilbert space, the action of symmetry transformations is captured
by one-body operators
\begin{equation}
\begin{split}
\hat{\Gamma}(\sym) \ket{ \varphi_{p} } &= 
\sum_{r} \Gamma(\sym)_{pr} \ket{ \varphi_{r} }  \quad, \\
\end{split}
\end{equation}
where $\Gamma(\sym)_{pr}$ is a $M$-dimensional matrix representation of $\mathcal{\symgroup}$.

\begin{figure*}[t!]
\centering
\includegraphics[width=0.8\textwidth]{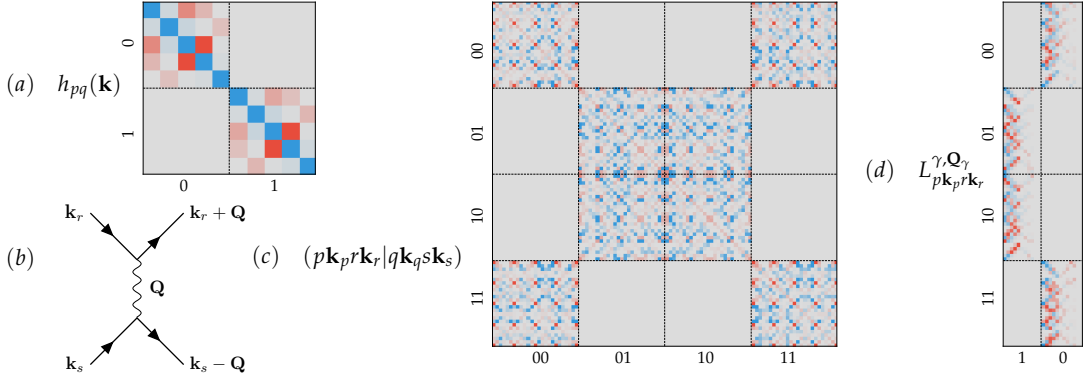}
\caption{(color online) Illustrative matrix elements of the Hamiltonian for a chain of 10 H atoms 
spaced by a distance of $R=1 \angstrom$, in the STO-6G basis.
To account for reflection symmetry across a plane perpendicular to the chain we construct a basis 
of orthonormal symmetry-adapted orbitals (even, odd denoted 0, 1 respectively).
(a) The one-body part of the Hamiltonian is block-diagonal.
(b) Conservation of $\kpt_r + \kpt_s$ (in this case parity) can be depicted in a two-vertex Feynman diagram.
(c) Sparsity in the electron repulsion integral (ERI), from the condition $\kpt_r + \kpt_s = \kpt_p + \kpt_q$.
(d) Cholesky decomposition of the ERI: $L^{\gamma \vett{Q}_\gamma}_{\kpt_p p , \kpt_r r} \neq 0$
only for $\vett{Q}_\gamma = \kpt_p - \kpt_r$.
Matrix elements are rescaled to enhance visibility, warm (cold) colors denote positive (negative) values.
}
\label{fig:sparseH}
\end{figure*}

To make the abstract group transformations $\oper{\sym}$ more concrete, and amenable 
to storage and numerical manipulation, it is useful to employ the structure theorem for finitely 
generated Abelian groups \cite{Hungerford_book_1980,Dummit_book_2004,Rudin_book_1962}, 
which states that $\mathcal{\symgroup}$ is isomorphic to a direct product 
\begin{equation}
\mathcal{\symgroup} \simeq 
\mathbb{Z}_{n_0} \times \dots \times \mathbb{Z}_{n_{r-1}} 
\equiv 
\mathbb{Z}_{\mathcal{\symgroup}}
\quad , \quad
\prod_{i=0}^{r-1} n_i = |\mathcal{\symgroup}|
\quad ,
\end{equation}
of cyclic groups $\mathbb{Z}_{n_i}$, of orders $n_i$ multiplying to the number 
$|\mathcal{\symgroup}|$ of elements of $\mathcal{\symgroup}$. 
In what follows,
symmetries $\hat{\sym} \in \mathcal{\symgroup}$ 
will be labeled with strings $\vett{\sym} \in \mathbb{Z}_{\mathcal{\symgroup}}$, and 
 sums of such strings will be understood to be modulo the orders $n_i$ of the cyclic groups,
\begin{equation}
\vett{\sym} + \vett{t} = ( s_0 + t_0 \mbox{ mod } n_0 \dots s_{r-1} + t_{r-1} \mbox{ mod } n_{r-1} ) \,\, .
\end{equation}
As detailed in Appendix \ref{sec:app_groups}, from the properties of the 
discrete Fourier transform \cite{Rudin_book_1962,Jozsa_RS_1998}, and the relation 
$\hat{\Gamma}(\vett{\sym}) \hat{\Gamma}(\vett{t}) = \hat{\Gamma}(\vett{\sym}+\vett{t})$, 
the operators
\begin{equation}
\hat{\Pi}_{\kpt} = \sum_{\vett{\sym}} 
\frac{e^{- 2\pi i \, \kpt \cdot \vett{\sym} }}{|\mathcal{\symgroup}|} 
\hat{\Gamma}(\vett{\sym})
\,\, , \,\,
\kpt \in \mathbb{Z}_{\mathcal{\symgroup}}
\,\, , \,\,
\kpt \cdot \vett{\sym} = \sum_{i=0}^{r-1} \frac{k_i s_i}{n_i}
\,\, ,
\end{equation}
form a complete set of orthogonal projectors,
\begin{equation}
\hat{\Pi}_{\kpt} \, \hat{\Pi}_{\kpt^\prime} = \delta_{\kpt \kpt^\prime} \, \hat{\Pi}_{\kpt} 
\quad,\quad
\sum_{\kpt} \hat{\Pi}_{\kpt} = \hat{ \mathbb{I} } \quad.
\end{equation}
Applying the projectors $\hat{\Pi}_{\kpt}$ to the basis functions, and orthonormalizing 
the resulting vectors,
yields an orthonormal basis of symmetry-adapted orbitals
$\{ \tilde{\varphi}_{p \kpt_p} \}$. Here, and in the remainder of the present work, 
$\kpt_p \in \mathbb{Z}_{\mathcal{\symgroup}}$ denotes an irreducible representation or irrep, 
and $p$ a group of $m_{\kpt_p}$ orbitals labelled by the irrep 
$\kpt_p$. The numbers $m_{\kpt_p}$ sum to $M$, and
\begin{equation}
\hat{\Pi}_{\kpt} \ket{\tilde{\varphi}_{p \kpt_p}} = 
\delta_{\kpt,\kpt_p} \, 
\ket{\tilde{\varphi}_{p \kpt_p}}
\,\, .
\end{equation}
The number of irreps will be denoted $N_k = |\mathcal{\symgroup}|$.
Since the orbitals $\tilde{\varphi}_{p \kpt_p}$ are eigenfunctions of the projectors 
$\hat{\Pi}_{\kpt}$, and the latter commute with the one- and two-body parts of the 
Hamiltonian, the matrix elements of $\oper{H}$ are sparse, as revealed 
by the expression
\begin{equation}
\label{eq:sparse_ham}
\begin{split}
&\oper{H} = E_0 + 
\sum_{ \substack{\kpt \\ pq} }  h_{pq}(\kpt) \crt{p \kpt} \dst{q \kpt} \\
+ & \sum_{ \substack{\kpt_p \kpt_r \kpt_q \kpt_s \\ prqs }}^*
\frac{(p \kpt_p , r \kpt_r | q \kpt_q , s \kpt_s)}{2} 
\crt{p \kpt_p} \crt{q \kpt_q} \dst{s \kpt_s} \dst{r \kpt_r}
\end{split}
\end{equation}
where the $*$ over the summation denotes the constraint $\kpt_p + \kpt_q = \kpt_r 
+ \kpt_s$ (with equality holding in the modular arithmetic of $\mathbb{Z}_{\mathcal{\symgroup}}$).
\eqref{eq:sparse_ham} can be rewritten in terms of the transfer parameter $\vett{Q} = \kpt_p - \kpt_r
= \kpt_s - \kpt_q$, leading to
\begin{equation}
\label{eq:sparse_ham2}
\begin{split}
\oper{H} &= E_0 + 
\sum_{ \substack{\kpt \\ pq} }  h_{pq}(\kpt) \crt{p \kpt} \dst{q \kpt} \\
&+ \sum_{ \substack{ \vett{Q} \kpt_r \kpt_s \\ prqs} }
\frac{(p \kpt_r + \vett{Q} , r\kpt_r | q \kpt_s - \vett{Q} , s\kpt_s)}{2} \,  \\
& \crt{p \kpt_r + \vett{Q}} \crt{q \kpt_s - \vett{Q}} \dst{s \kpt_s} \dst{r \kpt_r}
\end{split}
\end{equation}
The structure of the Hamiltonian operator is illustrated in Figure \ref{fig:sparseH}.
In the forthcoming sections, after providing a brief description of the AFQMC method, we will
discuss in detail how the form of the Hamiltonian in  \eqref{eq:sparse_ham2} leads to savings in
many of the operations in the method.

\subsection{The AFQMC method}

The AFQMC method \cite{Zhang_PRL90_2003,Purwanto_PRE70_2004,Zhang_Notes_2013,Motta_WIRES_2018} 
expresses the many-body ground state $\Psi_0$ of a Hamiltonian $\hat{H}$ through an imaginary time evolution,
\begin{equation}
|\Psi_0\rangle \propto \lim_{n\to\infty} e^{- n \, \Delta\tau (\hat{H}-E_0)} \, | \Phi_I \rangle
\quad ,
\end{equation}
where $\Delta\tau$ (the time step) is chosen small and the initial state $\Phi_I$, which 
should not be orthogonal to $\Psi_0$, is often a single Slater determinant.
To sample the many-body propagator, we rewrite 
the Hamiltonian as 
\begin{equation}
\hat{H} - E_0 = \hat{H}_1 - \frac{1}{2} \sum_\af \hat{v}_\af^2 \quad,
\label{eq:afqmc_friendly}
\end{equation}
where $\hat{H}_1$, $\hat{v}_\af$ are one-body operators.
Then, using a Hubbard-Stratonovich transformation \cite{Hubbard_PRL3_1959,Stratonovich_SPD2_1958}, the short-imaginary-time propagator is
\begin{equation}
e^{-\Delta\tau (\hat{H}-E_0) } = 
\int d{\bf{x}} \, p({\bf{x}}) \, \hat{B}({\bf{x}}) \, , \label{eq:shortprop}
\end{equation}
where 
\begin{equation}
\hat{B}({\bf{x}}) = e^{ - \frac{\Delta \tau}{2} \hat{H}_1 }
e^{ \sqrt{\Delta \tau} \sum_{\af} x_\af \hat{v}_\af }
e^{ - \frac{\Delta \tau}{2} \hat{H}_1 } 
\end{equation}
is a one-body  propagator that is a function of the multi-dimensional vector ${\bf{x}}$, and $p({\bf{x}})$ is the 
standard normal probability distribution. AFQMC thus represents the many-body wave function as a superposition
of non-orthogonal Slater determinants,
\begin{equation}
\label{eq:psi-PI-form}
\begin{split}
|\Phi_n \rangle 
& = e^{- n \, \Delta\tau (\hat{H}-E_0) } \ket{ \Phi_I }
= \int \prod_{l=0}^{n-1} d{\bf{x}}_l \,\, p({\bf{x}}_l) 
\hat{B}({\bf{x}}_l)  \ket{ \Phi_I } \, .
\end{split}
\end{equation}
The ground-state expectation value of $\hat{H}$ is obtained as
\begin{equation}
\begin{split}
\label{eq:esta}
&
\frac{\langle \Psi_T | \hat{H} | \Phi_n \rangle}{\langle \Psi_T | \Phi_n \rangle}
= \\ 
=
\, 
&\frac
{ \int \prod_{l=0}^{n-1} d{\bf{x}}_l \,\, p({\bf{x}}_l) \, \langle \Psi_T | \hat{H} | \prod_{l=0}^{n-1} \hat{B}({\bf{x}}_l)  \ket{ \Phi_I } }
{ \int \prod_{l=0}^{n-1} d{\bf{x}}_l \,\, p({\bf{x}}_l) \, \langle \Psi_T | \prod_{l=0}^{n-1} \hat{B}({\bf{x}}_l)  \ket{ \Phi_I } } \, = \\
=
\,
&\frac
{ \int d\vett{X} \, p(\vett{X}) \, W(\vett{X}) \, \mathcal{E}_{loc}( \Phi_n(\vett{X}) ) }
{ \int d\vett{X} \, p(\vett{X}) \, W(\vett{X}) } \quad , \\
\end{split}
\end{equation}
where $\Psi_T$ is a second many-body state, called the trial wavefunction.
In Eq.~\eqref{eq:esta}, the overlap and local energy,
\begin{equation}
\begin{split}
W(\vett{X}) &= \langle \Psi_T | \prod_{l=0}^{n-1} \hat{B}({\bf{x}}_l)  \ket{ \Psi_I } \equiv
\braket{ \Psi_T | \Phi_n(\vett{X}) }
\quad , \\
\mathcal{E}_{loc}( \Phi_n(\vett{X}) ) &= 
\frac{\langle \Psi_T | \hat{H} | \Phi_n(\vett{X}) \rangle }
{\langle \Psi_T | \Phi_n(\vett{X}) \rangle } \quad , \\
\end{split}
\end{equation}
are defined on a path $\vett{X} = ({\bf{x}}_{n-1} \dots {\bf{x}}_0)$ of auxiliary fields at each time slice
up to $n-1$. The expectation value is computed over a collection of Monte Carlo (MC) samples 
(labeled by $i$) as
\begin{equation}
\frac{
\langle \Psi_T | \hat{H} | \Psi_0 \rangle 
}{
\langle \Psi_T | \Psi_0 \rangle 
}
\simeq 
\frac{ \sum_{i} W_i \, \mathcal{E}_{loc}(\Phi_i) }
{ \sum_{i} W_i } \, .
\end{equation}
and the stochastically sampled determinants $\Phi_i$ are called walkers in the AFQMC literature.

Because the propagator $\hat{B}({\bf{x}})$ contains stochastically fluctuating fields, the MC sampling will lead to 
complex overlaps $W_i$, which causes the variance of this estimator to grow exponentially with the number of time-steps
$n$. This phase problem can be controlled by an approximate gauge condition known as
the phaseless approximation~\cite{Zhang_PRL90_2003,Purwanto_PRE70_2004,Zhang_Notes_2013,Motta_WIRES_2018}, which we summarize below:
\begin{enumerate}
\item mean-field background subtraction: \\
the expectation values
$\langle \hat{v}_\af
\rangle_{T} = \langle \Psi_T | \hat{v}_\af | \Psi_T \rangle$
are computed
and the Hamiltonian is 
rewritten as
\begin{equation}
\hat{H} - E_0 = \hat{H}^\prime_1 - \frac{1}{2} \sum_\af 
\left( \hat{v}_\af - \langle \hat{v}_\af \rangle_{T} \right)^2 \quad,
\label{eq:h_mfbg}
\end{equation}
\item importance sampling transformation: \\
the Hubbard-Stratonovich is 
defined up to a shift $\vett{x} \to \vett{x} - \overline{\vett{x}}$, $\overline{\vett{x}}
\in \mathbb{C}$, and this additional freedom is exploited to rewrite the estimator 
\eqref{eq:esta} with the replacements 

\begin{equation}
\begin{split}
&\hat{B}({\bf{x}}) \to 
\hat{B}^\prime(\vett{x}-\overline{\vett{x}}) = e^{ - \frac{\Delta \tau}{2} \hat{H}^\prime_1 }
e^{ \sqrt{\Delta \tau} \sum_{\af} (x_\af-\overline{x}_\af) \hat{v}^\prime_\af }
e^{ - \frac{\Delta \tau}{2} \hat{H}^\prime_1 } \,\, , \\
&\prod_{l=0}^{n-1} \hat{B}({\bf{x}}_l)  \ket{ \Phi_I }
\to
\prod_{l=0}^{n-1} \hat{B}^\prime(\vett{x}_l-\overline{\vett{x}}_l)  \ket{ \Phi_I } 
\equiv \ket{\Phi^\prime_n(\vett{X},\overline{\vett{X}})} \,\, , \\
&W^\prime(\vett{X}) \to \prod_{l=0}^{n-1} I\Big(\vett{x}_l,\overline{\vett{x}}_l;\Phi^\prime_l
(\vett{X},\overline{\vett{X}}) \Big) \,\, , \\
&I\big(\vett{x},\overline{\vett{x}};\Phi \big) = 
\frac{p(\vett{x}-\overline{\vett{x}})}{p(\vett{x})} 
\frac{ \langle \Psi_T | \hat{B}^\prime(\vett{x}_l-\overline{\vett{x}}_l)  \ket{ \Phi} }{
\langle \Psi_T \ket{ \Phi} } \,\, ,
\end{split}
\end{equation}
where $\hat{v}^\prime_\af = \hat{v}_\af - \langle \hat{v}_\af \rangle_{T}$. One makes the choice
\begin{equation}
\overline{\vett{x}}_\af = - \sqrt{\Delta \tau} \, 
\frac{ \langle \Psi_T | \hat{v}^\prime_\af \ket{ \Phi} }
{\langle \Psi_T \ket{ \Phi} } 
\equiv - \sqrt{\Delta \tau} \, 
\langle \hat{v}^\prime_\af \rangle
\end{equation}
to minimize fluctuations in the importance function $I(\vett{x},\overline{\vett{x}};
\Phi)$ to leading order in $\Delta\tau$.
\item Real local energy and cosine approximations: 
\\ the importance function is approximated as
\begin{equation}
\begin{split}
I\big(\vett{x},\overline{\vett{x}};\Phi \big) &\simeq
e^{ - \Delta\tau \, \mbox{Re} \, \left( \mathcal{E}_{loc}(\Phi) - E_0 \right) } \, 
\max\left( 0 , \cos( \Delta \theta) \right) \quad , \\
\Delta \theta &= \mbox{Arg} \, \frac{ \langle \Psi_T | \hat{B}^\prime(\vett{x}-\overline{\vett{x}}) \ket{ \Phi} }
{\langle \Psi_T \ket{ \Phi} } \quad .
\end{split}
\end{equation}
\end{enumerate}
Steps 1, 2 are simply re-parametrizations which reduce fluctuations in
the estimators of physical properties when $\overline{\bf{x}}_\gamma$ is real. When it is complex, these 
transformations ensure that the gauge variation is minimized, enabling to remove the sign problem in 
step 3. The bias resulting from the approximations in step 3 can be reduced by improving the trial wavefunction.
AFQMC has been successfully applied to multiple lattice models of correlated electrons
\cite{LeBlanc_PRX5_2015,Qin_PRB_2016,Zheng_Science_2017} and real materials 
\cite{Purwanto2014,Motta_PRX_2017,Shee_JCTC_2019} achieving accuracies competitive with
other high-level wavefunction methods.
There are many additional algorithmic improvements and extensions, for example
to compute properties~\cite{Motta_JCTC_2017,Motta_JCP_2017,Shee_JCTC13_2017},
and to improve the efficiency of the method~\cite{Motta_WIRES_2018,Motta_arxiv_2019}.

\section{Method}

We now illustrate
how to account for Hamiltonian symmetries in the above procedure.
Additional implementation details appear in the Appendices.
In what follows, we  consider  a symmetry group composed of $N_k$ Abelian symmetries, with irreps labelled 
by $\kpt$.
We  use the symbols $M$,$N$,$N_\af$ to denote the total number of basis functions, 
particles and auxiliary fields respectively. Correspondingly, we use
$m_\kpt$, $n_\kpt$ and  $n_{\afpi,\vett{k}}$ to denote the number of basis functions, 
particles and auxiliary fields labelled by the irrep $\kpt$, and $m$, $n$, $n_\af$ to denote
the average numbers of basis functions, particles and auxiliary fields per irrep, 
$m = \sum_\kpt m_\kpt/N_k$ etc.
Indices $prqs$, $ij$, $\afpi$ run over basis functions, particles and auxiliary fields labelled 
by a specific irrep $\kpt$, respectively.

\subsection{Hamiltonian representation} 

As expressed in \eqref{eq:afqmc_friendly}, in  AFQMC we must express
the two-body part of the Hamiltonian as a sum of squares of one-body operators.
To achieve this, one commonly relies on a density fitting (DF) \cite{DF1,DF2} or 
Cholesky decomposition (CD) 
\cite{Beebe_IJQC12_1977,Koch_JCP118_2003,Aquilante_JCC31_2010,Purwanto_JCP135_2011} 
of the electron repulsion integrals, 
$(pr|qs) \simeq \sum_{\af=1}^{N_\af} L_{pr}^\af L_{qs}^\af$ where $N_\af$ is
the number of components.
As detailed in Appendix \ref{sec:app}, in the presence of symmetries, such
a decomposition becomes
\begin{equation}
\frac{(p \kpt_r\vett{+Q}, r \kpt_r | q \kpt_s\vett{-Q}, s\kpt_s)}{2} = 
\sum_\afpi L^{\afpi,  {\vett{Q}}}_{p \kpt_r\vett{+Q} ,r\kpt_r}
L^{\afpi, - {\vett{Q}}}_{q \kpt_s\vett{-Q},s\kpt_s}
\, ,
\label{eq:CD}
\end{equation}
where, as illustrated in Figure \ref{fig:sparseH}, components $\afpi$ are labelled by irreps
 $\vett{Q}$, $-\vett{Q}$, the number of which $n_{\afpi,\vett{Q}}$ sums to $N_\af$. 
Then (as detailed in Appendix \ref{sec:work}) we interchange the creation and destruction 
operators in \eqref{eq:sparse_ham2} and use \eqref{eq:CD} to
rewrite the Hamiltonian as a sum of squares,
\begin{widetext}
\begin{equation}
\begin{split}
\oper{H} - E_0 &=  
\sum_{ \substack{\kpt \\ pq} } 
\tilde{h}_{pq}(\kpt)
\crt{p \kpt} \dst{q \kpt} 
- \frac{1}{2} \, \left[ 
\sum_{\afpi \vett{Q}}
\left( \frac{ i \hat{L}_{\afpi,\vett{Q}} + i \hat{L}_{\afpi,-\vett{Q}} }{\sqrt{2}} \right)^2 
+
\left( \frac{ \hat{L}_{\afpi,\vett{Q}} - \hat{L}_{\afpi,-\vett{Q}} }{\sqrt{2}} \right)^2 \right]
\quad,
\label{eq:almost_there}
\end{split}
\end{equation}
\end{widetext}
where 
\begin{equation}
\label{eq:lvecs}
\hat{L}_{\afpi,\vett{Q}}
= \sum_{\substack{\kpt_r \\ rp}} 
L^{\afpi,{\vett{Q}}}_{p \kpt_r\vett{+Q} ,r\kpt_r}  \crt{p \kpt_r + \vett{Q}} \dst{r \kpt_r}  \quad .
\end{equation}
The total number of auxiliary fields is thus $2 N_\af$,  where the correspondence with the label $\af$ in
\eqref{eq:afqmc_friendly} is
 $\af \to (\afpi,\vett{Q},1)$, $(\afpi,\vett{Q},2)$ and the auxiliary field operators are
 $\hat{v}_{\afpi,\vett{Q},1} = (i \hat{L}_{\afpi,\vett{Q}} + i \hat{L}_{\afpi,-\vett{Q}} ) / 
\sqrt{2}$ and $\hat{v}_{\afpi,\vett{Q},2} = ( \hat{L}_{\afpi,\vett{Q}} - \hat{L}_{\afpi,-\vett{Q}} ) / 
\sqrt{2}$ respectively.
A simple technique to reduce the number of auxiliary fields to $N_\af$ 
is described in Appendix \ref{sec:lagrange}.

Importantly, the correction to the one-body part of the Hamiltonian resulting from
the interchange of creation and destruction operators in \eqref{eq:sparse_ham2}
does not mix orbitals labeled by different irreps (i.e. it is block-diagonal).
Additional details are given in Appendix \ref{app:anticomm}.

It is clear that the Hamiltonian integrals, such as $(p \kpt_r\vett{+Q}, r \kpt_r | q \kpt_s\vett{-Q}, s\kpt_s)$, 
require $1/N_k$
less storage than without symmetry. The coefficients  defining $\hat{L}_{\gamma,\vett{Q}}$ also show
a $1/N_k$ reduction in storage compared to without symmetry.

\subsection{Mean-field wavefunction and background subtraction} 

We assume that the trial wavefunction is a single determinant.
If the single determinant wavefunction  transforms as an irrep
of $\mathcal{S}$, then its orbitals  $\psi_{i {\kpt_i}}$ can also be labelled by irreps, thus
\begin{equation}
\label{eq:sparse_psit}
\begin{split}
\ket{\Psi_T} &= \prod_{i \kpt_i} \crt{\psi_{i \kpt_i}} \ket{\emptyset} \quad , \\
\crt{\psi_{i \kpt_i}} &= \sum_{r} \Big( {\Psi_T}(\kpt_i) \Big)_{ri} \, \crt{r \kpt_r} \quad , \\
\end{split}
\end{equation}
where the numbers $n_\kpt$ of particles in each orbital irrep satisfy  
$\sum_{\kpt} n_\kpt = N$
and the coefficient matrix ${\Psi_T}(\kpt_i)$ is blocked by symmetry.
In contrast, a generic Slater determinant (such as the walkers  in an AFQMC calculation)
is parametrized by a dense $M\times N$ matrix $\Phi_{r \kpt_r , i \kpt_i}$.
An important property of
\eqref{eq:sparse_psit} is that the one-body density matrix
\begin{equation} 
\begin{split}
\rho_{r \kpt_r , p \kpt_p} &= 
\braket{ \Psi_T |\crt{p \kpt_p} \dst{r \kpt_r} | \Psi_T } \\
&=  \delta_{\kpt_p \kpt_r} 
\left[ \Psi_T(\kpt_r) \Psi_T^\dagger(\kpt_r) \right]_{rp} 
\end{split}
\end{equation}
is also block-diagonal.
The mean-field expectation values of the operators $\hat{L}_{\afpi,\vett{Q}}$ 
thus read
\begin{equation} 
\label{eq:ell}
\begin{split}
\langle \Psi_T | \hat{L}_{\afpi,\vett{Q}} | \Psi_T \rangle 
= 
\delta_{\vett{Q},\vett{0}} \, \ell_{\afpi}
\,\, .
\end{split}
\end{equation}
Defining the operators
\begin{equation}
\oper{L}^\prime_{\afpi,\vett{Q}} = 
\oper{L}_{\afpi,\vett{Q}} - \, \frac{\delta_{\vett{Q},\vett{0}} \, \ell_\afpi}{N} \, \oper{N} \quad,
\end{equation}
which by construction have zero average over $\Psi_T$,
we can obtain operators with the mean-field background subtracted, as in
\eqref{eq:h_mfbg} and detailed in Appendix \ref{sec:mfbg},
\begin{equation}
\begin{split}
\hat{v}^\prime_{\afpi\vett{Q},1} &= 
\frac{ i \hat{L}^\prime_{\afpi,\vett{Q}} + i \hat{L}^\prime_{\afpi,-\vett{Q}} }{\sqrt{2}} \\
\hat{v}^\prime_{\afpi\vett{Q},2} &= 
\frac{ \phantom{i} \hat{L}^\prime_{\afpi,\vett{Q}} - \phantom{i} \hat{L}^\prime_{\afpi,-\vett{Q}} }{\sqrt{2}}  \\
\end{split}
\end{equation}
In this step, the trial wavefunction requires storing $\sum_\kpt m_\kpt n_\kpt \simeq mn N_k$ coefficients, 
a reduction of $1/N_k$ compared to without symmetry. Similarly, the block structure of the trial wavefunction 
and density matrix means that computing the mean-field density matrix and subsequent background term
also involves a $1/N_k$ reduction in the number of operations.

\subsection{Overlap calculation} 
\label{sec:ovlp}

The overlap matrix between the trial wavefunction and a walker
\begin{equation}
\ket{\Phi} = \prod_{j \kpt_j} \crt{\Phi_{j \kpt_j} } \ket{\emptyset}
\end{equation}
is $\braket{\Psi_T| \Phi} = \det(\Omega)$, with
\begin{equation}
\begin{split}
\left( \Omega \right)_{i \kpt_i , j \kpt_j} 
= \sum_r \Big( \Psi_T(\kpt_i) \Big)^\dagger_{ir} \Phi_{r \kpt_i , j \kpt_j}
\quad .
\end{split}
\end{equation}
Due to the symmetry block structure of $\Psi_T$, computing $\Omega$ requires $\mathcal{O}(m N^2)$ 
operations, even though $\Omega$ is in general  dense.
Compared to the corresponding cost of $\mathcal{O}(MN^2)$ in a calculation that does not exploit 
symmetry, this is more efficient by a factor of $1/N_k$.

\subsection{Force bias calculation} 

Using symmetry, the force bias calculation involves the quantities $\langle \hat{v}^\prime_\af \rangle$.
These are easily related to
\begin{equation} 
\langle \hat{L}_{\afpi,\vett{Q}} \rangle
\equiv \sum_{ \substack{\kpt_r \\ pr } } L^{\afpi,\vett{Q}}_{p \kpt_p,r\kpt_r} 
\frac{\langle \Psi_T | \crt{p \kpt_p} \dst{r \kpt_r} | \Phi \rangle}
{ \langle \Psi_T | \Phi \rangle }
\quad,
\end{equation}
with ${\vett{Q}} = \kpt_p - \kpt_r$, which are calculated as
\begin{equation}
\begin{split}
\langle \hat{L}_{\afpi,\vett{Q}} \rangle &= 
\sum_{ \substack{ \kpt_r \\ ir} }
\mathcal{L}^{\afpi,\vett{Q}}_{i \kpt_p , r \kpt_r} 
\Theta_{ r \kpt_r , i \kpt_p }
\quad , \\
\mathcal{L}^{\afpi,\vett{Q}}_{i \kpt_p, r \kpt_r} &= \sum_p
\Big( \Psi_T(\kpt_p) \Big)^\dagger_{ip} \, L^{\afpi, \vett{Q} }_{p \kpt_p , r \kpt_r} \quad ,
\end{split}
\end{equation}
where $\vett{Q} = \vett{k}_p - \kpt_r$
and $\Theta = \Omega^{-1} \Phi$. $\mathcal{L}$ can be precomputed, storing
$m n n_\af N_k^2$ complex numbers. 
The computational cost to obtain the force bias with precomputation is
$\mathcal{O}(mn n_\af N_k^2)$.
Compared with the cost of a calculation that does not exploit symmetry ($MN N_\af$ operations) 
this is more efficient by a factor $1/N_k$.
Note, however, that while we can use symmetry when generating the matrix $\Omega$, as
discussed in Section \ref{sec:ovlp}, the cost of computing its determinant and its 
inverse is not reduced by symmetry, because it is in general dense. There is thus only a partial 
gain due to symmetry for the force bias calculation.

\subsection{Walker propagation} 

With symmetry, the small-imaginary time propagator in \eqref{eq:shortprop}
takes the form
\begin{equation}
\label{eq:aofx}
\begin{split}
e^{-\Delta \tau (\oper{H}-E_0)} = e^{ - \frac{\Delta \tau}{2} \oper{H}_1^\prime }
\int 
d \vett{x} \, e^{ \hat{A}(\vett{x}) }
e^{ - \frac{\Delta \tau}{2} \oper{H}_1^\prime }
\quad ,
\end{split}
\end{equation}
for 
operators $\hat{A}(\vett{x})$, defined as a linear combination of 
$x_{\vett{Q}\gamma 1}\hat{v}^\prime_{\gamma,\vett{Q},1}$,
$x_{\vett{Q}\gamma 2}\hat{v}^\prime_{\gamma,\vett{Q},2}$,
as derived in Appendix \ref{sec:lagrange}.
Applying the exponential of $\hat{H}^\prime_1$ to a (walker) Slater determinant can be
carried out efficiently. In fact, since $\hat{H}^\prime_1$ is symmetric and
thus does not mix
irreps, it has the form
$\hat{H}^\prime_1 = \sum_{ \kpt pq } h^\prime_{pq}(\kpt) \,
\crt{p \kpt} \dst{q \kpt}$. Thus if the walker $\ket{\Phi}$ is 
parametrized by the matrix $\Phi_{p \kpt_p,i \kpt_i}$, its image $\ket{\Phi^\prime} 
= e^{ - \frac{\Delta \tau}{2} \oper{H}_1^\prime } \ket{\Phi}$ from the one-body 
propagator is parametrized by the matrix
\begin{equation}
\left( \Phi^\prime \right)_{p \kpt_p , i \kpt_i} 
= 
\sum_{q} \Big( e^{ - \frac{\Delta \tau}{2} h^\prime(\kpt_p)} \Big)_{pq}
\big( \Phi^\prime \big)_{q \kpt_p , i \kpt_i}  \quad .
\end{equation}
This can be computed using $\mathcal{O}(m^2 N N_k)$ operations, as compared to $\mathcal{O}(M^2 N)$ operations 
in a calculation that does not use symmetry.
To propagate the interacting part of $\oper{H}$, one possible strategy is to
construct the matrix $\mathcal{A}_{p \kpt_p, r \kpt_r}$ associated with the
operator $\hat{A}(\vett{x})$ and then apply $e^{\mathcal{A}}$ to $\Phi$.
The cost to construct $\mathcal{A}$ is $\mathcal{O}(m^2 n_\gamma N_k^2)$ 
(it is a linear combination of block sparse matrices), a reduction of $1/N_k$ compared to without symmetry.
However, since in general $\mathcal{A}$ lacks any sparsity properties, the cost of applying its exponential 
to $\Phi$ requires $\mathcal{O}(M^2 N)$ operations.
Note that, in principle, sparsity
could be exploited 
by applying individual terms of $e^{\hat{A}}$, such as $e^{\sqrt{\Delta\tau} i x_{\vett{Q}\gamma 1}\hat{v}^\prime_{\gamma,\vett{Q},1}}$, where such an
exponential is further expanded as a power series with each term transforming as an irrep and corresponding to sparse matrix multiplication onto $\Phi$. 
However, this is only a savings for symmetry groups where the number of irreps far exceeds the number of terms in the power series.
We have not observed savings with this second strategy in this work. Thus we only report calculations where we apply the full $\mathcal{A}$ matrix, 
with only a partial (i.e. limited to the generation of $\mathcal{A}$) reduction in cost in this step from symmetry.

\begin{table*}[t!]
\begin{tabular}{llll}
\hline \hline
operation & without symmetry & with symmetry & savings \\
\hline
storage of integrals & $M^2+M^2N_\af+MN$ & $m^2 N_k + m^2 n_\af N_k^2 + mn N_k$ & $N_k^{-1}$ \\
no. of auxiliary fields & $N_\af$ & $n_\af N_k$ & none \\
\hline
force bias calculation & $MN N_\af$ & $mn n_\af N_k$ & $N_k^{-1}$ \\
overlap, $\Theta$ matrix & $MN^2+N^3 + M^2 N$ & $m N^2 + N^3 + M^2 N$ & $N_k^{-1}$ (partial) \\
\hline
propagation (kinetic) & $M^2 N$ & $m^2 N N_k$ & $N_k^{-1}$ \\
propagation (potential) & $M^2 N_\af + M^2 N$ & $m^2 n_\af N_k + M^2 N$ & $N_k^{-1}$ (partial) \\
\hline
local energy calculation & $NM+M N^2 N_\af$ & $mn N_k +m n^2 n_\af N_k^3$ & $N_k^{-1}$  \\
\hline \hline
\end{tabular}
\caption{Comparison of the computational cost of the main operations in an AFQMC calculation, without
(left) and with the use of symmetry (right). In most cases, a cost reduction by a factor of $N_k^{-1}$ is seen.
The three terms under ``storage of integrals'' refer to the one- and two-body parts of $\oper{H}$ and the trial
wavefunction respectively. The two terms under ``overlap, $\Theta$ matrix'' refer to the construction and 
inversion of the overlap matrix and construction of the $\Theta$ matrix respectively. The two terms 
under ``propagation (potential)'' refer to construction and application of the $\mathcal{A}$ matrix.
The two terms under ``local energy calculation'' refer to one- and two-body parts of the local energy.}\label{tab:sym}
\end{table*}

\subsection{Local energy calculation} 

For the one-body part of the local energy, the form of the estimator and the speedup over 
implementations without symmetry are very similar to the case of the force bias. Indeed, 
an analogous calculation leads to the expressions
\begin{equation}
\begin{split}
\mathcal{E}_{loc,1}(\Phi) = 
\sum_{ \substack{\kpt_i \\ ir } } \mathcal{K}_{ir}(\kpt_i) \Theta_{r \kpt_i , i \kpt_i}
\, , \\
\mathcal{K}_{ir}(\kpt) = \sum_p \Big( \Psi_T(\kpt) \Big)^\dagger_{ip} \, h_{pr}(\kpt) \\ 
\end{split}
\end{equation}
For the considerably more expensive two-body part, we use the generalized Wick's theorem
\cite{Wick_PR80_1950,Balian_NC64_1969} to obtain 
\begin{equation}
\begin{split}
\mathcal{E}_{loc,2}(\Phi) = \sum_{\vett{Q} \afpi} \sum_{ \substack{ \kpt_r \kpt_s \\ ij } }
f^\afpi_{ i \kpt_p , i \kpt_r } f^\afpi_{ j \kpt_q , j \kpt_s } 
- 
f^\afpi_{ i \kpt_p , j \kpt_q } f^\afpi_{ j \kpt_q , i \kpt_p } 
\end{split}
\end{equation}
where $ij$ are associated with particles labelled by the irreps $\kpt_p = \kpt_r + \vett{Q}$, 
$\kpt_q = \kpt_s - \vett{Q}$ respectively, and the tensor $f$ is defined as 
\begin{equation}
f^{\afpi}_{i \kpt_p , j \kpt_q } = 
\sum_{r} \mathcal{L}^{\afpi}_{i \kpt_p , r \kpt_r} \Theta_{r \kpt_r, j \kpt_q}
\quad .
\end{equation}
As seen, the cost of the procedure is $\mathcal{O}(m n^2 n_\af N_k^3)$ operations to generate the 
tensor $f$, and $\mathcal{O}(n^2 n_\af N_k^3)$ to perform the final contraction.
Compared with a calculation without symmetry, this is more efficient by a factor of $1/N_k$.

\subsection{Summary} 

The acceleration achieved in AFQMC due to the use of symmetries is summarized in Table \ref{tab:sym}.
Most computational steps are accelerated by a factor of $1/N_k$, and storage is reduced by $1/N_k$ as well. 
In a standard mean-field calculation, symmetries lead to an acceleration by a factor of $1/N_k^2$, due
to symmetries in both the Hamiltonian as well as the wavefunction. In AFQMC, the acceleration is
limited to $1/N_k$ because the walkers do not transform as irreps of $\mathcal{S}$, as individual
components of the Hubbard-Stratonovich such as $\hat{L}_{\gamma,\vett{Q}}$ all transform as different irreps 
of $\mathcal{S}$.

\section{Results}
\label{sec:results}

We now present some illustrative calculations using symmetry in AFQMC calculations for a molecule
and for crystalline
systems.
Restricted Hartree-Fock (RHF), density functional theory (DFT), M{\o}ller-Plesset perturbation theory (MP2) 
and coupled-cluster with singles and doubles (CCSD) calculations were performed with the PySCF package
\cite{Sun_WIRES_2018}.

Molecular calculations were all-electron calculations using the cc-pVTZ basis~\cite{dunning1989cc1,dunning1989cc3}.
The auxiliary field decomposition was performed using Cholesky decomposition, and the RHF state was used
as a trial wavefunction in the AFQMC calculations.

In the crystal calculations presented below, core electrons were replaced by  norm-conserving GTH Pad\'e 
pseudopotentials \cite{Goedecker_PRB_1996,Hartwigsen_PRB_1998,Berkelbach_JCTC_2017}.
Hamiltonian matrix elements were computed with the PySCF program~\cite{Sun_WIRES_2018} using the 
GTH series of Gaussian bases~\cite{cp2kbasisref}.
Gaussian density fitting was used to treat the electron-electron interaction and
to obtain the auxiliary field decomposition~\cite{sun2017gaussian}.
RHF energies are reported using
the leading finite size correction for the $\vett{G}=0$ contribution to the Hartree-Fock exchange (\texttt{exxdiv=ewald})
and total energies from other methods were obtained by adding this RHF energy to the respective correlation energies, with integrals computed
omitting the $\vett{G}=0$ term (\texttt{exxdiv=None})~\cite{Berkelbach_JCTC_2017,Sun_WIRES_2018}.
The RHF state was used as a trial wavefunction in the AFQMC calculations.

\subsection{Molecular systems and point group symmetry}

As a simple test-case, we first consider a molecular system, SF$_6$, where we use the reflection group symmetry.
The reflection group is isomorphic to $\mathbb{Z}_2^3$, where $3$ is the number of reflection planes, giving
8 irreps in the group. 

\begin{figure}[h!]
\centering
\includegraphics[width=0.45\textwidth]{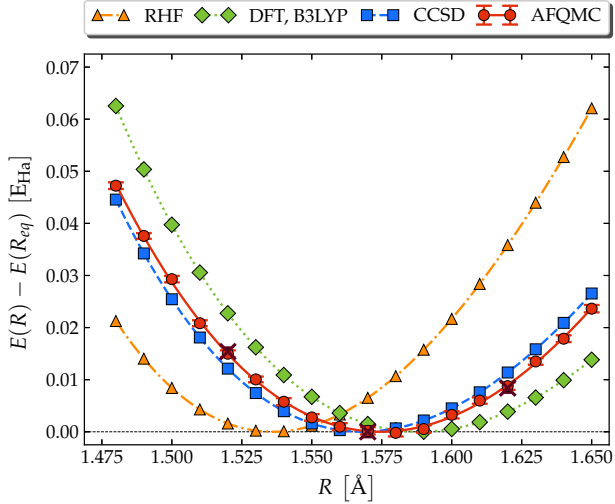}
\caption{
Energy as function of bondlength for SF$_6$ for RHF, DFT-B3LYP, CCSD and AFQMC 
with and without reflection symmetry (red circles, dark red crosses) in the cc-pVTZ basis. 
Energies are shown relative to the minimum value $E(R_{eq})$.}
\label{fig:mols}
\end{figure}

In Fig.~\ref{fig:mols} we show the equation of state $E(R)$ of the molecule (where $R$ is the S-F bond length) using 
AFQMC, DFT with the B3LYP functional, RHF and CCSD.
As can be seen, the AFQMC calculations performed with and without reflection symmetry (red circles
and dark red crosses in Fig.~\ref{fig:mols}) yield identical results to within statistical error. CCSD and AFQMC yield 
potential energy surfaces in good agreement with each other.

\subsection{Crystalline solids}

The computational saving from symmetries is especially important in systems with a large symmetry group, and 
crystalline solids form one such example. Consider a crystal with a primitive cell with lattice vectors 
$\vett{a}_0$, $\vett{a}_1$, $\vett{a}_2$. We use translational-symmetry-adapted (crystalline)  
Gaussian atomic orbitals \cite{Berkelbach_JCTC_2017} (AOs) as a symmetry basis.
Starting from a set of Gaussian AOs in the primitive cell $\varphi_{\mu}$,
these can be written as
\begin{equation}
\ket{ \tilde{\varphi}_{\mu \vett{k}} } 
= 
\sum_{\vett{i}} \frac{ e^{- 2 \pi i \, \vett{i} \cdot \vett{k}} }{\sqrt{N_k}} 
\, 
\ket{ \varphi_{\mu \vett{i}} } \quad ,
\end{equation}
where $\vett{i}$ is an integer vector $(i_0 , i_1 , i_2)$ 
denoting $\varphi_\mu$ translated from the primitive cell by lattice vector 
$\sum_{r=0}^2 i_r \vett{a}_r$, and $\vett{k}$ has the form $\sum_{r=0}^2 \frac{ k_r }{N_r} \vett{b}_r$,
where $k_r$ is an integer vector with $0 \leq k_r < N_r$ and
$\vett{b}_r$ are the reciprocal lattice vectors. This choice of $k_r$ is equivalent to sampling 
the Brillouin zone with a mesh of $N_0 \times N_1 \times N_2$ wave-vectors including the $\Gamma$ point (origin).
Note the above basis representation spans the same Hilbert space as a $N_0 \times N_1 \times N_2$ supercell
calculation with ($\Gamma$ point) periodic boundary conditions~\cite{Evarestov_book}.

The Hamiltonian symmetries are the lattice translations, corresponding to integer multiples $\sum_{r=0}^2 s_r \vett{a}_r$ of 
the $\vett{a}_0$, $\vett{a}_1$, $\vett{a}_2$ vectors. Under such translations, the basis transforms as
\begin{align}
\hat{\Gamma}(\vett{\sym}) \ket{ \tilde{\varphi}_{\mu \vett{k}} } = 
e^{2 \pi i \sum_r \frac{s_r k_r}{N_r}} \ket{ \tilde{\varphi}_{\mu \vett{k}} }
\end{align}
thus, the translation group is isomorphic to $\mathbb{Z}_{N_0} \times \mathbb{Z}_{N_1} \times \mathbb{Z}_{N_2}$.

\begin{figure}[h!]
\includegraphics[width=0.45\textwidth]{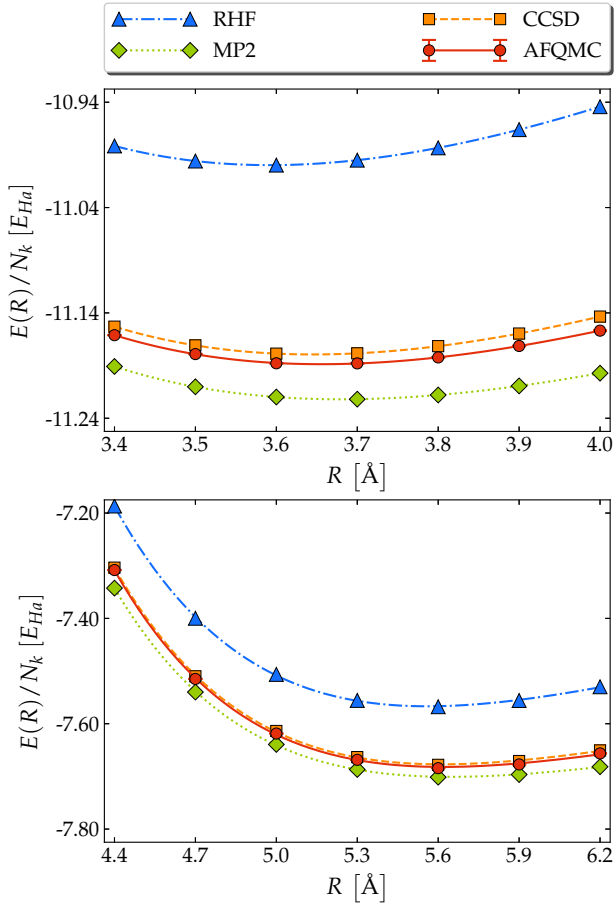}
\caption{
Equation of state of C diamond (top) and Si FCC (bottom), using a 2$\times$2$\times$2 $k$-point mesh and the 
GTH-DZV basis and GTH Pad\'e pseudopotential, from RHF, MP2, CCSD, AFQMC (blue triangles, green diamonds, orange squares, red circles).
\label{fig:eb1}
}
\end{figure}

To demonstrate the symmetry-adapted AFQMC using the lattice translation group we first
compute the equilibrium lattice constants of C diamond and Si FCC in Figure \ref{fig:eb1}, using
a 2$\times$2$\times$2 $k$-point mesh, at the GTH-DZV level. Here we find that AFQMC is in good agreement with
CCSD using the same $k$-point mesh, and significantly improves on RHF and MP2.
This trend can be seen in the potential energy surfaces in Figure \ref{fig:eb1}, as well as in the
corresponding equilibrium lattice constants.

Using translational symmetry, we can further consider larger symmetry groups in order
to extrapolate to the thermodynamic limit (TDL). Note that increasing the size of
the translational symmetry group yields the same result as a calculation with increased supercell size,
but with much reduced cost.

\begin{figure}[h!]
\centering
\includegraphics[width=0.5\textwidth]{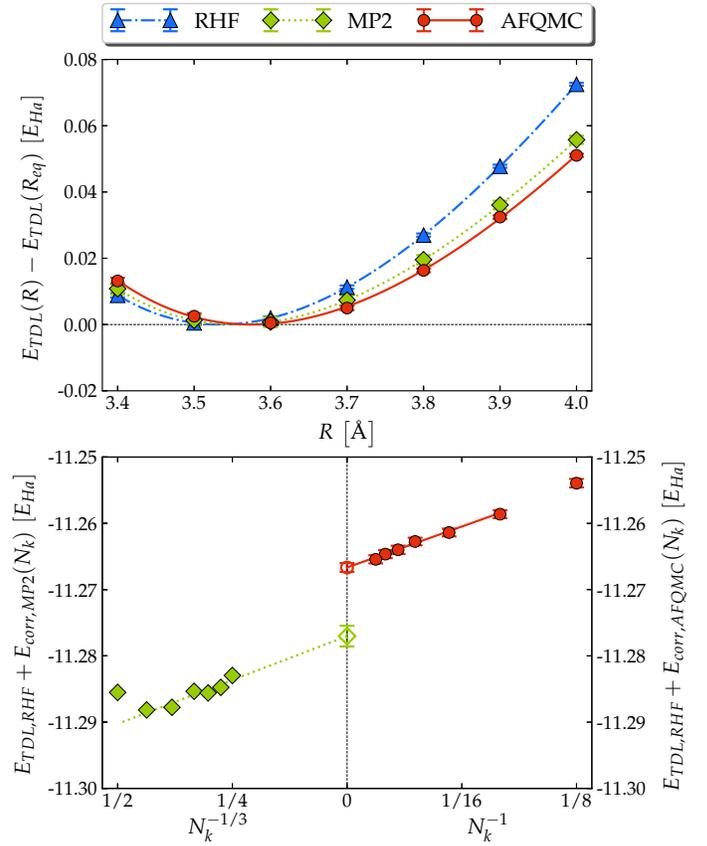}
\caption{
Top: Equation of state of C diamond from RHF, MP2 and AFQMC (blue triangles, green diamonds, red circles)
extrapolated to the thermodynamic limit, in the GTH-DZV basis, using the GTH-Pad\'{e} pseudopotential.
Bottom: detail of the thermodynamic limit extrapolation for the MP2 (left) and AFQMC (right) correlation energy.
Extrapolated quantities are shown with empty symbols.
}
\label{fig:tdl1}
\end{figure}

We illustrate the extrapolation of results to the thermodynamic limit in Figure \ref{fig:tdl1}, using
C diamond as a test system.
RHF and correlation energies were computed for 2$\times$2$\times$2, 2$\times$2$\times$3, 3$\times$3$\times$3,
3$\times$3$\times$4, 4$\times$4$\times$3 and 4$\times$4$\times$4 meshes of $k$-points, using the GTH-DZV basis.
In the upper panel of Figure \ref{fig:tdl1} we show the equation of state $E_{TDL}(R)$ 
extrapolated to the thermodynamic limit (with the minimum value $E_{TDL}(R_{eq})$ subtracted) 
from RHF, MP2 and AFQMC.

We extrapolate RHF total energies and AFQMC, MP2 correlation energies (per cell) to the TDL using
power-law Ansatz $E(N_k) = \alpha + \beta N_k^\mu$, with $\mu = - 1$ for RHF and AFQMC \cite{Kwee_PRL100_2008}
and $\mu = - \frac{1}{3}$ for MP2 \cite{Berkelbach_JCTC_2017}.
Extrapolation of RHF (correlation) energies is carried out using data for all but the smallest two (the 
smallest) $k$-point meshes.
In the lower panel of Figure \ref{fig:tdl1}, we illustrate the extrapolation of MP2 (left) and AFQMC (right)
correlation energies at the representative bondlength $R=3.6 \, \angstrom$.

Fitting the TDL curves to the Morse potential Ansatz $E(R) = E_0 + \Delta E \, ( 1 - e^{-\alpha (R-R_{eq})} )^2$ gives
an equilibrium bondlength of $R_{eq,AFQMC} = 3.575(1) \, \angstrom$. For the 2$\times$2$\times$2 supercell,
the same procedure yields $R_{eq,AFQMC} = 3.657(1) \, \angstrom$, thus TDL extrapolation significantly 
shortens the AFQMC equilibrium bondlength.
For reference, the experimental bondlength is $R = 3.553 \, \angstrom$, corrected for zero-point vibrational effects 
\cite{Schima_JCP_2011}; for a more faithful comparison, a larger basis set should be used \cite{Morales_arxiv_2018a}.

\begin{figure}[h!]
\centering
\includegraphics[width=0.45\textwidth]{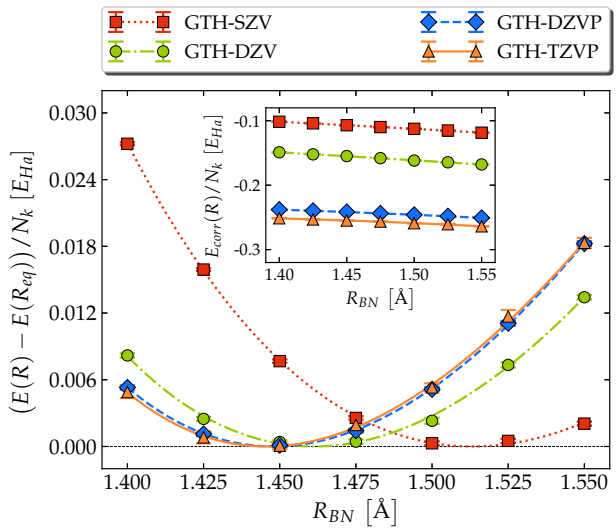}
\caption{
AFQMC total (main figure) and correlation energy per cell (inset) of 2D hexagonal BN,
for increasingly large basis sets, using a 4$\times$4$\times$1 $k$-point mesh.
Total energies are shown relative to the minimum value, attained at equilibrium bondlength, $E(R_{eq})$.
}
\label{fig:tdl2}
\end{figure}

In Figure \ref{fig:tdl2} we carry out a similar calculation for 2D
hexagonal boron nitride. RHF energies and AFQMC correlation energies (inset) were computed for 4$\times$4$\times$1
meshes of $k$-points.
Total energies are shown in the upper panel, measured from the minimum value, $E_{min} = 
E(R_{eq})$. The AFQMC equilibrium bondlengths are $R_{eq} = 1.5133(9)$, $1.4613(16)$, $1.4478(5)$, $1.4455(12)$ $\angstrom$
for GTH-SZV, GTH-DZV, GTH-DZVP and GTH-TZVP respectively; for reference, the reported
experimental equilibrium bondlength is $R_{eq}=1.45 \angstrom$ \footnote{\,https://github.com/cryos/avogadro/blob/master/\\crystals/nitrides/BN.cif}.

So far,  we have illustrated the use of symmetry when calculating total energies and lattice constants.
We now briefly show that symmetry adaptation can be used when computing arbitrary ground-state
properties in AFQMC, such as the electron density, within the back-propagation algorithm
\cite{Zhang_PRB55_1997,Purwanto_PRE70_2004,Motta_JCTC_2017}.

The electron density is computed by contracting the spin-summed one-body density matrix
with the basis orbitals $\varphi_{p \kpt_p}({\bf{x}})$, evaluated on a mesh of points ${\bf{x}}$ along 
the lattice plane,
\begin{equation}
\rho({\bf{x}}) = \sum_{{\substack{\kpt_p \kpt_q \\ pq \, \sigma }}}
\varphi^*_{p \kpt_p}({\bf{x}}) \varphi_{q \kpt_q}({\bf{x}}) \, \rho^\sigma_{p \kpt_p , q \kpt_q} \quad .
\end{equation}
The one-body density matrix is evaluated using the back-propagation algorithm as
\begin{equation}
\rho^\sigma_{p \kpt_p , q \kpt_q} = \frac{1}{\sum_i W_i } \,
\sum_{i} W_i 
\frac{\langle \Psi_i | \hat{a}^\dag_{p \kpt_p \sigma} \hat{a}^{\phantom{\dag}}_{q \kpt_q \sigma} 
| \Phi_i \rangle}{ \langle \Psi_i | \Phi_i \rangle } \quad.
\end{equation}
Here, $|\Psi_i \rangle$ is a stochastically sampled Slater determinant sampled
at imaginary time $n \, \Delta \tau$, $W_i$ its future weight at some time $(n+m)
\, \Delta \tau$ and $\langle \Phi_i|$ is obtained back-propagating (i.e. propagating 
as a bra or linear functional, rather than a ket or vector) $\langle \Psi_T|$ along 
the segment of the future path of $| \Psi_i \rangle$, sampled during the time 
interval between $n \, \Delta \tau$ and $(n+m) \, \Delta \tau$ \cite{Motta_JCTC_2017}.

\begin{figure}[h!]
\centering
\includegraphics[width=0.45\textwidth]{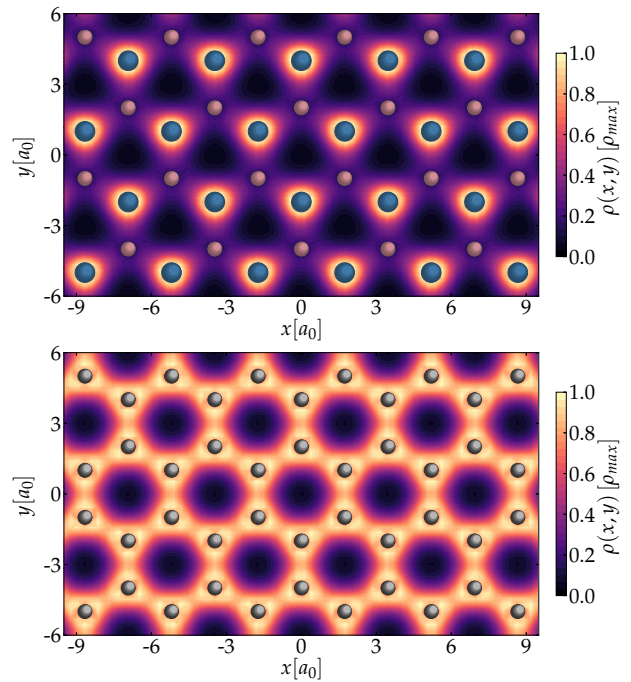}
\caption{
AFQMC ground-state density of 2D hexagonal BN (top) and graphene (bottom) at the experimental
equilibrium lattice constant $a_0$, using the GTH-DZV basis and GTH-Pad\'{e} pseudopotential, 
along the lattice plane. Pink small (gray small, large blue) spheres denote B (C, N) atoms.
}
\label{fig:rho}
\end{figure}

In Figure \ref{fig:rho}, 
we compute the electronic density of two low-dimensional materials, 2D hexagonal BN and graphene, 
within the GTH-DZV basis. The electron density illustrates the different nature of the two materials: while 
in graphene the  density is distributed uniformly around C atoms, in BN there is a net concentration of 
electrons around N atoms, consistent with the charge-transfer nature of the material. 

\subsection{Timings}
\begin{figure*}[t!]
\centering
\includegraphics[width=0.8\textwidth]{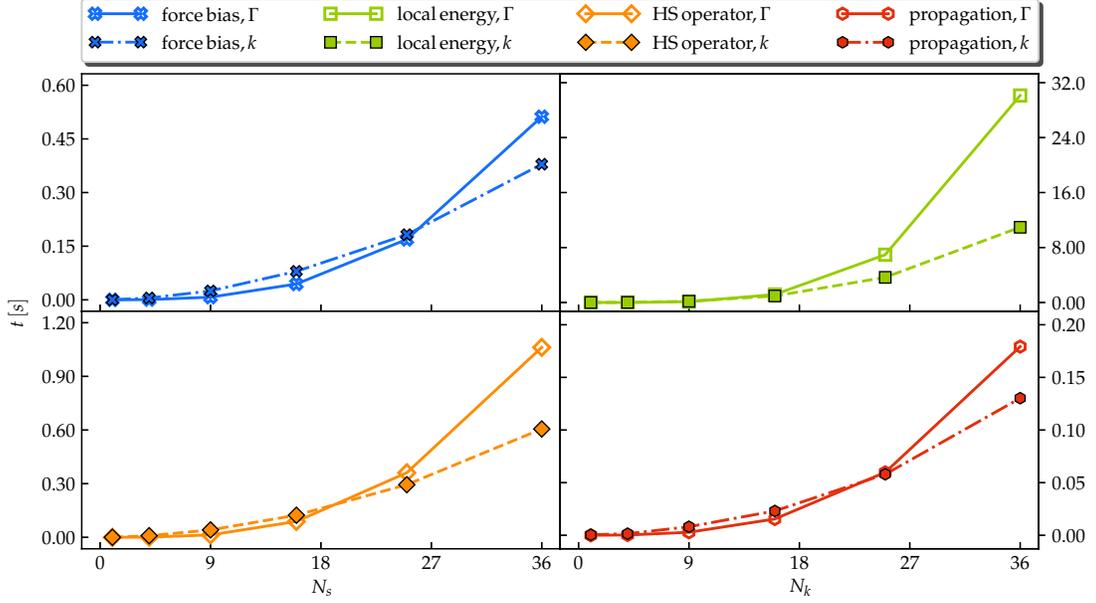}
\caption{
(color online) Left to right and top to bottom: force bias (blue crosses), local energy (green squares), 
Hubbard-Stratonovich operator construction (orange diamonds) and walker propagation (red hexagons) 
times as a function of supercell size $N_s$ (solid lines, empty symbols) or $k$-point mesh size $N_k$ 
(dashed lines, filled symbols) for BN using the GTH-DZV basis and GTH-Pad\'e pseudopotential.
}
\label{fig:tdl3}
\end{figure*}

In Figure \ref{fig:tdl3} we compare the timings of the standard and symmetry-adapted AFQMC 
implementations. Timings were performed on a cluster with Intel E5-2680, 2.4 GHz CPUs.
In the various panels, the times for force bias and local energy evaluation, Hubbard-Stratonovich 
operator construction and walker propagation, the most expensive steps of an AFQMC calculation,
are shown for standard AFQMC calculations of BN at the GTH-DZV level, using supercells of
increasingly large size $N_s$, and symmetry-adapted calculations using $k$-point meshes of 
increasingly large size $N_k$.
In a standard calculation, the local energy evaluation scales as $N_s^4$ and all the other subroutines
as $N_s^3$.
In a symmetry-adapted calculation, the local energy evaluation scales as $N_k^3$ and all the other 
subroutines scale as $N_k^2$, confirming the reduction in scaling by one power arising from symmetry
adaptation.
In the current implementation, the lower scaling comes at the cost of an increased prefactor, so 
that crossover between the two strategies occurs around $N_s = N_k \simeq 10$ for the local
energy evaluation and $N_s = N_k \simeq 20$ for all other subroutines. 

\section{Conclusions}
\label{sec:conclusions}

In this work, we presented a formalism to perform AFQMC calculations that 
take advantage of Abelian Hamiltonian symmetries. We described how within
a symmetry adapted orbital basis, the matrix elements of the Hamiltonian operator acquire
block sparsity, which, when combined with a trial state that transforms
as an irrep of the symmetry group and Hubbard-Stratonovich fields that
also transform as irreps of the symmetry group, it is possible
to reduce the cost and memory of the main steps in the AFQMC calculation by a factor of $N_k^{-1}$,
where $N_k$ is the order of the group.

Extending this formalism to non-Abelian symmetries is straightforward.
The only difference arises because irreps of non-Abelian groups need not be one-dimensional. Thus products
of objects that transform as irreps (such as $\varphi^*_{p \vett{k}_p}(\vett{r}) \, \varphi_{r \vett{k}_r}(\vett{r})$) no longer simply transform
as a single irrep $\vett{k}_r - \vett{k}_p$, but correspond to a linear combination of objects,
each transforming according to potentially different irreps.
Nonetheless, all quantities that are block sparse in the current algorithm remain
block sparse in the non-Abelian generalization, and a similar speedup of $\mathcal{O}(N_k^{-1})$ will be achieved
as it is observed here.

As we showed in our demonstration calculations, the use of Abelian symmetries is particularly
beneficial in the context of the large translational group associated with crystalline calculations. Thus we believe
the present work will be particularly important in accelerating AFQMC calculations in realistic materials,
and in particular, in removing finite size effects and in extrapolations to the thermodynamic limit.

\section{Acknowledgments}

M. M. acknowledges Qiming Sun and James McClain for assistance and discussions regarding ES calculations for crystalline solids.
This work was supported by the US Department of Energy, Office of Science (via Grant No. SC0019390 to
G. K.-L. C.).
S. Z. acknowledges support from DOE (Grant No. DE-SC0001303).
Additional software developments for Hamiltonian symmetries implemented in PySCF were supported by US NSF (Grant No. 1657286).
Computations were carried out on facilities supported by the US Department of Energy, National Energy Research Scientific Computing Center (NERSC), 
on facilities supported by the Scientific Computing Core at the Flatiron Institute, on the Pauling cluster at the California Institute of Technology, and on the 
Storm and SciClone Clusters at the College of William and Mary. The Flatiron Institute is a division of the Simons Foundation.

\appendix

\section{Additional theoretical details}

\subsection{Properties of the $\hat{\Pi}_\kpt$ operators}
\label{sec:app_groups}

The relation
\begin{equation}
\sum_{\vett{k}} 
\frac{e^{- 2\pi i \, \kpt \cdot \vett{\sym} }}{|\mathcal{\symgroup}|} 
= 
\prod_{i=0}^{r-1} \sum_{k_i=0}^{N_i-1} 
\frac{e^{- 2\pi i \, \frac{k_i \sym_i}{n_i} }}{n_i}
= 
\delta_{\vett{\sym},\vett{0}}
\end{equation}
readily implies that the operators $\hat{\Pi}_\kpt$ are orthogonal projectors,
\begin{equation}
\begin{split}
\hat{\Pi}_\kpt \hat{\Pi}_{\kpt^\prime} 
&= 
\sum_{\vett{\sym} \vett{\sym}^\prime} 
\frac{ e^{- 2\pi i \, (\kpt \cdot \vett{\sym}-\kpt^\prime \cdot \vett{\sym}^\prime) } }
{ |\mathcal{\symgroup}|^2 }
\hat{\Gamma}(\vett{\sym}) \hat{\Gamma}(\vett{\sym}^\prime) 
= \\
&= \sum_{\vett{\sym} \vett{t}} 
\frac{ e^{- 2\pi i \, \big( \kpt \cdot \vett{\sym}-\kpt^\prime \cdot (\vett{t}-\vett{\sym}) \big) } }
{ |\mathcal{\symgroup}|^2 }
\hat{\Gamma}(\vett{t}) = \\
&= \sum_{\vett{\sym}} 
\frac{ e^{- 2\pi i \, \big( (\kpt -\kpt^\prime) \cdot \vett{\sym} \big) } }
{ |\mathcal{\symgroup}| } 
\hat{\Pi}_{\kpt^\prime}
= \delta_{\kpt , \kpt^\prime} \hat{\Pi}_{\kpt^\prime} \quad .
\end{split}
\end{equation}
Completeness holds, since
\begin{equation}
\sum_\kpt \hat{\Pi}_\kpt = 
\sum_{\vett{\sym} \kpt} \frac{ e^{- 2\pi i \, \kpt \cdot \vett{\sym} }}
{ |\mathcal{\symgroup}| } 
\hat{\Gamma}(\vett{\sym}) 
= \sum_{\vett{\sym}} 
\delta_{\vett{\sym},\vett{0}} 
\hat{\Gamma}(\vett{\sym}) = \hat{\mathbb{I}} \quad ,
\end{equation}
and the neutral element $\vett{0}$ of $\mathbb{Z}_{\mathcal{S}}$ is mapped onto
the neutral element $\hat{\Gamma}(\vett{0}) = \hat{\mathbb{I}}$ of $\mathcal{S}$.
Finally,
\begin{equation}
\begin{split}
\hat{\Gamma}(\vett{\sym}) \hat{\Pi}_\kpt = 
\sum_{\vett{t}} 
\frac{e^{- 2\pi i \, \kpt \cdot \vett{t} }}{|\mathcal{\symgroup}|} 
\hat{\Gamma}(\vett{\sym}+\vett{t}) 
= 
e^{2\pi i \, \kpt \cdot \vett{\sym} } \, \hat{\Pi}_\kpt \quad .
\end{split}
\end{equation}
Hamiltonian sparsity easily follows from the fact that symmetry-adapted orbitals are
eigenfunctions of projection operators, and that projection operators commute with
the one-body and two-body parts of the Hamiltonian. Indeed,
\begin{equation}
\begin{split}
&
\langle p \vett{k}_p | \hat{H}_1 | q \vett{k}_q \rangle =
\langle p \vett{k}_p | \hat{\Pi}_{\vett{k}_p} \hat{H}_1 | q \vett{k}_q \rangle = \\
= \,
& 
\langle p \vett{k}_p | \hat{H}_1 \hat{\Pi}_{\vett{k}_p} | q \vett{k}_q \rangle
=
\delta_{\vett{k}_p \vett{k}_q} h_{pq}(\kpt)
\end{split}
\end{equation}
and, since the projectors onto symmetry adapted orbitals in the two-particle Hilbert space become
\begin{equation}
\hat{\Pi}^{(2)}_\kpt = \sum_{\kpt_1} \hat{\Pi}_{\kpt_1} \otimes \hat{\Pi}_{\kpt-\kpt_1} \quad ,
\end{equation}
one has
\begin{equation}
\begin{split}
&\phantom{=} ( p \vett{k}_p r \vett{k}_r | q \vett{k}_q s \vett{k}_s ) = 
\langle p \vett{k}_p q \vett{k}_q | \hat{H}_2 | r \vett{k}_r s \vett{k}_s \rangle = \\
&= \langle p \vett{k}_p q \vett{k}_q | \Pi^{(2)}_{\kpt_p + \kpt_q} \hat{H}_2 | r \vett{k}_r s \vett{k}_s \rangle = \\ 
&= \langle p \vett{k}_p q \vett{k}_q | \hat{H}_2 \Pi^{(2)}_{\kpt_p + \kpt_q} | r \vett{k}_r s \vett{k}_s \rangle = \\
&= \delta_{\kpt_p + \kpt_q , \kpt_r + \kpt_s} ( p \vett{k}_p r \vett{k}_r | q \vett{k}_q s \vett{k}_s )
\quad .
\end{split}
\end{equation}

\subsection{Density fitting and Cholesky decomposition}
\label{sec:app}

In this Section we show how the structure \eqref{eq:CD} emerges when the electron-electron
interaction is treated within the density fitting (DF) or Cholesky (CD) decomposition.
Within DF, the electron repulsion integral is approximated by density fitting with an auxiliary basis 
of atom-centered Gaussian atomic orbitals $\{ \chi_\af \}_{\af=1}^{N_\af}$,
\begin{equation}
(pr|qs) \simeq \sum_{\af \afB} (pr|\af) S^{-1}_{\af \afB} (\afB |qs) \quad .
\end{equation}
where $S_{\gamma\delta} = \langle \chi_\gamma | \chi_\delta \rangle$. The action of the symmetry 
group on the auxiliary basis $\{\chi_\af \}_\af$ is captured by a family of operators
\begin{equation}
\hat{\Gamma}(\vett{\sym}) | \chi_\af \rangle = \sum_\delta \Gamma(s)_{\af \afB} | \chi_\afB \rangle \quad ,
\end{equation}
so that, by following the procedure outlined in Section \ref{sec:sparse}, one can produce an
orthonormal basis of symmetry-adapted auxiliary basis functions $\tilde{\chi}_{\afpi \vett{Q}}$
(with overlap matrix equal to the identity). The electron repulsion integral reads, in the 
symmetry-adapted molecular and auxiliary bases,
\begin{equation}
\frac{(p \kpt_r + \vett{Q} , r \kpt_r | q \kpt_s - \vett{Q} ,s  \kpt_s )}{2} = \sum_{\afpi}
L^{\afpi,\vett{Q}}_{p \kpt_r + \vett{Q} , r \kpt_r}
L^{\afpi,-\vett{Q}}_{q \kpt_s - \vett{Q} ,s  \kpt_s}
\end{equation}
where summation is restricted to auxiliary basis functions belonging to the irrep
labelled by $\vett{Q}$ for the pair $(p \kpt_r + \vett{Q} , r \kpt_r)$ and by $-\vett{Q}$ for the pair
$(q\kpt_s - \vett{Q}  , s\kpt_s)$ respectively.

Performing a Cholesky decomposition of the electron repulsion integral,
\begin{equation}
\frac{(p \vett{k}_p r \vett{k}_r | q \vett{k}_q s \vett{k}_s)}{2} 
= 
\sum_{\af} L^\af_{p \kpt_p , r \kpt_r} L^\af_{q \kpt_q , s \kpt_s}
\end{equation}
may not lead to the form \eqref{eq:CD}, i.e. the tensor $L$ may not be sparse. The desired 
structure can be extracted performing a SVD of the rank-three tensor $L^\af_{q \kpt_q , s \kpt_s} = 
\sum_\mu U^\mu_{q \kpt_q , s \kpt_s} \sigma_\mu V^{\mu \af}$
After the SVD is taken, the ERI reads
\begin{equation}
\frac{(p \vett{k}_p r \vett{k}_r | q \vett{k}_q s \vett{k}_s)}{2} 
= 
\sum_{\mu} U^\mu_{ p \kpt_p , r \kpt_r } \sigma_\mu^2 U^\mu_{ q \kpt_q , s \kpt_s } \quad ,
\end{equation}
and the tensor $U$ is non-zero only for certain values of the index $\mu$, that depend only 
on the difference $\kpt_p - \kpt_r = \vett{Q}$. Indices $\mu$ can thus be parametrized as 
pairs $\afpi,\vett{Q}$, and the ERI takes the desired form in Sec. \ref{sec:sparse}.

\section{Additional algorithmic details}
\label{sec:work}

\subsection{Interaction as squares of one-body operators}
\label{app:anticomm}

Starting from \eqref{eq:sparse_ham2}, we interchange the creation and destruction operators,
\begin{equation}
\begin{split}
   &\crt{p \kpt_r + \vett{Q}} \crt{q \kpt_s - \vett{Q}} \dst{s \kpt_s} \dst{r \kpt_r} =
   - \crt{p \kpt_r + \vett{Q}} \crt{q \kpt_s - \vett{Q}} \dst{r \kpt_r} \dst{s \kpt_s} = \\
= \, &\crt{p \kpt_r + \vett{Q}} \dst{r \kpt_r} \crt{q \kpt_s - \vett{Q}} \dst{s \kpt_s} 
   - \delta_{r \kpt_r,q \kpt_s - \vett{Q}} \crt{p \kpt_r + \vett{Q}} \dst{s \kpt_s} \\
\end{split}
\end{equation}
and inserting this equation in \eqref{eq:sparse_ham2}, obtaining
\begin{equation}
\begin{split}
&\oper{H} - E_0 =  \\
&\sum_\kpt \left( h_{pq}(\kpt) - \sum_{\kpt_r r} 
\frac{(p \kpt  ,r  \kpt_r |  r \kpt_r , q \kpt )}{2} \right)
\crt{p \kpt} \dst{q \kpt} \\
&+ 
\sum_{\afpi \vett{Q}} 
\hat{L}_{\afpi,\vett{Q}} \hat{L}_{\afpi,-\vett{Q}} \\
\end{split}
\end{equation}
with $\hat{L}_{\afpi,\vett{Q}}$ as in Eq. \eqref{eq:lvecs}.
To obtain a representation as a sum of squares of one-body operators, we observe that
\begin{equation}
\begin{split}
&\sum_{\vett{Q} \afpi}
\hat{L}_{\afpi,\vett{Q}} \hat{L}_{\afpi,-\vett{Q}}
=
\frac{1}{2} 
\sum_{\vett{Q} \afpi}
\hat{L}_{\afpi,\vett{Q}} \hat{L}_{\afpi,-\vett{Q}}
+
\hat{L}_{\afpi,-\vett{Q}} \hat{L}_{\afpi,\vett{Q}} = \\
&- \frac{1}{2} \left[ \sum_{\vett{Q} \afpi}
\left( \frac{ i \hat{L}_{\afpi,\vett{Q}}  + i \hat{L}_{\afpi,-\vett{Q}} }{\sqrt{2}} \right)^2 
+
\left( \frac{ \hat{L}_{\afpi,\vett{Q}} - \hat{L}_{\afpi,-\vett{Q}} }{\sqrt{2}} \right)^2 \right] \, .
\end{split}
\end{equation}

\subsection{Mean-field background subtraction}
\label{sec:mfbg}

The mean-field background subtraction requires replacing the operators $\oper{L}_{\vett{Q}, \afpi}$
with $\oper{L}^\prime_{\vett{Q}, \afpi}$ in \eqref{eq:almost_there}. This leads to 
\begin{equation}
\label{eq:mfbg_intermediate}
\begin{split}
&\oper{H} - E_0 = \sum_\afpi \ell_\afpi^2 + 2 \sum_\afpi \ell_\afpi \oper{L}^\prime_{\vett{0},\afpi} \\
+ \, &\sum_{ \substack{\kpt pq \\ \sigma} } 
\left( h_{pq}(\kpt) -
\frac{1}{2} \, \sum_{\kpt_r r} 
(p\kpt , r \kpt_r | r \kpt_r , q\kpt ) \right)
\crt{p \kpt} \dst{q \kpt}  \\
- \, & \frac{1}{2} \, \left[ 
\sum_{\vett{Q} \afpi}
\left( \frac{ i \hat{L}^\prime_{\afpi,\vett{Q}} + i \hat{L}^\prime_{\afpi,-\vett{Q}} }{\sqrt{2}} \right)^2 
+
\left( \frac{ \hat{L}^\prime_{\afpi,\vett{Q}} - \hat{L}^\prime_{\afpi,-\vett{Q}} }{\sqrt{2}} \right)^2 \right]
\quad ,
\end{split}
\end{equation}
where $\ell$ is defined as in Eq. \eqref{eq:ell}.
The operator \eqref{eq:mfbg_intermediate} has the same form as in \eqref{eq:h_mfbg} with 
$\hat{H}^\prime_1 = \sum_\kpt h^\prime_{pq}(\kpt) \crt{p \kpt} \dst{q \kpt} $,
\begin{equation}
\begin{split}
h^\prime_{pq}(\kpt) &= 
h_{pq}(\kpt) - \frac{1}{2} \, \sum_{\kpt_r r} (p\kpt , r \kpt_r | r \kpt_r , q\kpt) + \\
&+ 2 \sum_\afpi \ell_\afpi L^{\afpi, \vett{0}}_{ p \kpt , q \kpt}  
- \frac{\ell_\afpi^2}{N} \delta_{pq} \quad ,
\end{split}
\end{equation}
and $\hat{v}^\prime_\af$ as detailed in the main text.

\subsection{Reducing the number of auxiliary fields \\ by Lagrangian partition}
\label{sec:lagrange}

In \eqref{eq:almost_there}, the Hamiltonian was expressed as
\begin{equation}
\begin{split}
&\oper{H} = E_0 + \oper{H}_1^\prime \\
&- \frac{1}{2} \, \left[
\sum_{\vett{Q} \afpi}
\left( \frac{ i \hat{L}^\prime_{\afpi,\vett{Q}} + i \hat{L}^\prime_{\afpi, -\vett{Q}} }{\sqrt{2}} \right)^2 
+
\left( \frac{ \hat{L}^\prime_{\afpi,\vett{Q}} - \hat{L}^\prime_{\afpi,-\vett{Q}} }{\sqrt{2}} \right)^2 \right]
\quad,
\end{split}
\end{equation}
clearly leading to $2N_\gamma$ auxiliary fields. This is more than in a calculation that
does not incorporate symmetry. To reduce the number of auxiliary fields,
we partition the irreps $\vett{Q}$ of the symmetry group
into three sets: 
\begin{itemize}
\item the set $\mathcal{P}_0 = \{ \vett{Q} : \vett{Q} = -\vett{Q} \}$ of $\vett{Q}$ 
coinciding with their inverse
\item any subset $\mathcal{P}_+ \subset \mathbb{Z}_\mathcal{\symgroup} - \mathcal{P}_0$
such that, if $\vett{Q} \in \mathcal{P}_+$, then $- \vett{Q} \notin \mathcal{P}_+$
\item $\mathcal{P}_- = 
\mathbb{Z}_\mathcal{\symgroup} - \mathcal{P}_0 - \mathcal{P}_+$
\end{itemize}
According to Lagrange's theorem \cite{Hungerford_book_1980,Dummit_book_2004,Rudin_book_1962}, 
the set $\mathcal{P}_0$ contains elements other than $\vett{0}$ in, and only in, groups with 
even order $|\mathcal{\symgroup}|$. Then clearly one has
\begin{equation}
\begin{split}
&\oper{H} = E_0 + \oper{H}_1^\prime 
- \frac{1}{2} \sum_{\vett{Q} \in \mathcal{P}_0, \afpi} 
\left( \sqrt{2} \oper{L}^\prime_{\vett{Q}, \afpi} \right)^2  \\
&- \frac{1}{2} \left[ \sum_{\vett{Q} \in \mathcal{P}_+, \afpi} 
\left( i \oper{L}^\prime_{\afpi,\vett{Q}} + i \oper{L}^\prime_{\afpi,-\vett{Q}} \right)^2
+
\left( \oper{L}^\prime_{\afpi,\vett{Q}} - \oper{L}^\prime_{\afpi,-\vett{Q}} \right)^2 \right] \quad .
\end{split}
\end{equation}
Now, since $|\mathcal{P}_0| + 2 |\mathcal{P}_+| = |\mathcal{\symgroup}|$, the interaction part of the
Hamiltonian has been reduced to a sum of $N_\gamma$ squares of one-body operators, the same 
as in a calculation that does not  enforce symmetries.

With this representation of the Hamiltonian, the small-imaginary-time propagator has the form
\eqref{eq:aofx} with 
\begin{equation}
\begin{split}
\int 
&d \vett{x} \, e^{ \hat{A}(\vett{x}) }=
\int 
\prod_{\vett{Q} \in \mathcal{P}_0 \afpi} 
dx_{\vett{Q} \afpi 1}
\prod_{\vett{Q} \in \mathcal{P}_+ \afpi} 
dx_{\vett{Q} \afpi 1} dx_{\vett{Q} \afpi_2} \\
&e^{ \sqrt{2 \Delta\tau} \sum_{\afpi,\vett{Q} \in \mathcal{S}} i \,
x_{\vett{Q} \afpi 1}\oper{L}^\prime_{\afpi,\vett{Q}} } \cdot \\
&e^{ \sqrt{\Delta\tau }
\sum_{\afpi,\vett{Q} \in \mathcal{P}_+}
i x_{\vett{Q} \afpi 1} 
\left( \oper{L}^\prime_{\afpi,\vett{Q}} + \oper{L}^\prime_{\afpi,-\vett{Q}} \right) } \\
&e^{ \sqrt{\Delta\tau }
\sum_{\afpi,\vett{Q} \in \mathcal{P}_+} 
x_{\vett{Q} \afpi 2} 
\left( \oper{L}^\prime_{\afpi,\vett{Q}} - \oper{L}^\prime_{\afpi,-\vett{Q}} \right)
}
\quad .
\end{split}
\end{equation}
This leads immediately to the form of the matrix $\mathcal{A}$ associated with $\hat{A}(\vett{x})$.


\begin{thebibliography}{59}%
\makeatletter
\providecommand \@ifxundefined [1]{%
 \@ifx{#1\undefined}
}%
\providecommand \@ifnum [1]{%
 \ifnum #1\expandafter \@firstoftwo
 \else \expandafter \@secondoftwo
 \fi
}%
\providecommand \@ifx [1]{%
 \ifx #1\expandafter \@firstoftwo
 \else \expandafter \@secondoftwo
 \fi
}%
\providecommand \natexlab [1]{#1}%
\providecommand \enquote  [1]{``#1''}%
\providecommand \bibnamefont  [1]{#1}%
\providecommand \bibfnamefont [1]{#1}%
\providecommand \citenamefont [1]{#1}%
\providecommand \href@noop [0]{\@secondoftwo}%
\providecommand \href [0]{\begingroup \@sanitize@url \@href}%
\providecommand \@href[1]{\@@startlink{#1}\@@href}%
\providecommand \@@href[1]{\endgroup#1\@@endlink}%
\providecommand \@sanitize@url [0]{\catcode `\\12\catcode `\$12\catcode
  `\&12\catcode `\#12\catcode `\^12\catcode `\_12\catcode `\%12\relax}%
\providecommand \@@startlink[1]{}%
\providecommand \@@endlink[0]{}%
\providecommand \url  [0]{\begingroup\@sanitize@url \@url }%
\providecommand \@url [1]{\endgroup\@href {#1}{\urlprefix }}%
\providecommand \urlprefix  [0]{URL }%
\providecommand \Eprint [0]{\href }%
\providecommand \doibase [0]{http://dx.doi.org/}%
\providecommand \selectlanguage [0]{\@gobble}%
\providecommand \bibinfo  [0]{\@secondoftwo}%
\providecommand \bibfield  [0]{\@secondoftwo}%
\providecommand \translation [1]{[#1]}%
\providecommand \BibitemOpen [0]{}%
\providecommand \bibitemStop [0]{}%
\providecommand \bibitemNoStop [0]{.\EOS\space}%
\providecommand \EOS [0]{\spacefactor3000\relax}%
\providecommand \BibitemShut  [1]{\csname bibitem#1\endcsname}%
\let\auto@bib@innerbib\@empty
\bibitem [{\citenamefont {{Born}}\ and\ \citenamefont
  {{Oppenheimer}}(1927)}]{Born_1927}%
  \BibitemOpen
  \bibfield  {author} {\bibinfo {author} {\bibfnamefont {M.}~\bibnamefont
  {{Born}}}\ and\ \bibinfo {author} {\bibfnamefont {R.}~\bibnamefont
  {{Oppenheimer}}},\ }\href {\doibase 10.1002/andp.19273892002} {\bibfield
  {journal} {\bibinfo  {journal} {Annalen der Physik}\ }\textbf {\bibinfo
  {volume} {389}},\ \bibinfo {pages} {457} (\bibinfo {year}
  {1927})}\BibitemShut {NoStop}%
\bibitem [{\citenamefont {Ziman}(1965)}]{Ziman_book_1965}%
  \BibitemOpen
  \bibfield  {author} {\bibinfo {author} {\bibfnamefont {J.}~\bibnamefont
  {Ziman}},\ }\href@noop {} {\emph {\bibinfo {title} {Principles of the theory
  of solids}}}\ (\bibinfo  {publisher} {University Press},\ \bibinfo {year}
  {1965})\BibitemShut {NoStop}%
\bibitem [{\citenamefont {Szabo}\ and\ \citenamefont
  {Ostlund}(1989)}]{Szabo_book_1989}%
  \BibitemOpen
  \bibfield  {author} {\bibinfo {author} {\bibfnamefont {A.}~\bibnamefont
  {Szabo}}\ and\ \bibinfo {author} {\bibfnamefont {N.}~\bibnamefont
  {Ostlund}},\ }\href@noop {} {\emph {\bibinfo {title} {Modern Quantum
  Chemistry: Introduction to Advanced Electronic Structure Theory}}},\ Dover
  Books on Chemistry\ (\bibinfo  {publisher} {Dover Publications},\ \bibinfo
  {year} {1989})\BibitemShut {NoStop}%
\bibitem [{\citenamefont {Dupuis}\ and\ \citenamefont
  {King}(1977)}]{dupuis1977molecular}%
  \BibitemOpen
  \bibfield  {author} {\bibinfo {author} {\bibfnamefont {M.}~\bibnamefont
  {Dupuis}}\ and\ \bibinfo {author} {\bibfnamefont {H.~F.}\ \bibnamefont
  {King}},\ }\href {\doibase 10.1002/qua.560110408} {\bibfield  {journal}
  {\bibinfo  {journal} {International Journal of Quantum Chemistry}\ }\textbf
  {\bibinfo {volume} {11}},\ \bibinfo {pages} {613} (\bibinfo {year}
  {1977})}\BibitemShut {NoStop}%
\bibitem [{\citenamefont {Stanton}\ \emph {et~al.}(1991)\citenamefont
  {Stanton}, \citenamefont {Gauss}, \citenamefont {Watts},\ and\ \citenamefont
  {Bartlett}}]{Stanton_JCP_1991}%
  \BibitemOpen
  \bibfield  {author} {\bibinfo {author} {\bibfnamefont {J.~F.}\ \bibnamefont
  {Stanton}}, \bibinfo {author} {\bibfnamefont {J.}~\bibnamefont {Gauss}},
  \bibinfo {author} {\bibfnamefont {J.~D.}\ \bibnamefont {Watts}}, \ and\
  \bibinfo {author} {\bibfnamefont {R.~J.}\ \bibnamefont {Bartlett}},\ }\href
  {\doibase 10.1063/1.460620} {\bibfield  {journal} {\bibinfo  {journal} {The
  Journal of Chemical Physics}\ }\textbf {\bibinfo {volume} {94}},\ \bibinfo
  {pages} {4334} (\bibinfo {year} {1991})}\BibitemShut {NoStop}%
\bibitem [{\citenamefont {Chan}\ and\ \citenamefont
  {Head-Gordon}(2002)}]{chan2002highly}%
  \BibitemOpen
  \bibfield  {author} {\bibinfo {author} {\bibfnamefont {G.~K.-L.}\
  \bibnamefont {Chan}}\ and\ \bibinfo {author} {\bibfnamefont {M.}~\bibnamefont
  {Head-Gordon}},\ }\href {\doibase 10.1063/1.1449459} {\bibfield  {journal}
  {\bibinfo  {journal} {The Journal of Chemical Physics}\ }\textbf {\bibinfo
  {volume} {116}},\ \bibinfo {pages} {4462} (\bibinfo {year}
  {2002})}\BibitemShut {NoStop}%
\bibitem [{\citenamefont {McClain}\ \emph {et~al.}(2017)\citenamefont
  {McClain}, \citenamefont {Sun}, \citenamefont {Chan},\ and\ \citenamefont
  {Berkelbach}}]{Berkelbach_JCTC_2017}%
  \BibitemOpen
  \bibfield  {author} {\bibinfo {author} {\bibfnamefont {J.}~\bibnamefont
  {McClain}}, \bibinfo {author} {\bibfnamefont {Q.}~\bibnamefont {Sun}},
  \bibinfo {author} {\bibfnamefont {G.~K.-L.}\ \bibnamefont {Chan}}, \ and\
  \bibinfo {author} {\bibfnamefont {T.~C.}\ \bibnamefont {Berkelbach}},\ }\href
  {\doibase 10.1021/acs.jctc.7b00049} {\bibfield  {journal} {\bibinfo
  {journal} {Journal of Chemical Theory and Computation}\ }\textbf {\bibinfo
  {volume} {13}},\ \bibinfo {pages} {1209} (\bibinfo {year} {2017})},\ \bibinfo
  {note} {pMID: 28218843}\BibitemShut {NoStop}%
\bibitem [{\citenamefont {Sun}\ \emph {et~al.}()\citenamefont {Sun},
  \citenamefont {Berkelbach}, \citenamefont {Blunt}, \citenamefont {Booth},
  \citenamefont {Guo}, \citenamefont {Li}, \citenamefont {Liu}, \citenamefont
  {McClain}, \citenamefont {Sayfutyarova}, \citenamefont {Sharma},
  \citenamefont {Wouters},\ and\ \citenamefont {Chan}}]{Sun_WIRES_2018}%
  \BibitemOpen
  \bibfield  {author} {\bibinfo {author} {\bibfnamefont {Q.}~\bibnamefont
  {Sun}}, \bibinfo {author} {\bibfnamefont {T.~C.}\ \bibnamefont {Berkelbach}},
  \bibinfo {author} {\bibfnamefont {N.~S.}\ \bibnamefont {Blunt}}, \bibinfo
  {author} {\bibfnamefont {G.~H.}\ \bibnamefont {Booth}}, \bibinfo {author}
  {\bibfnamefont {S.}~\bibnamefont {Guo}}, \bibinfo {author} {\bibfnamefont
  {Z.}~\bibnamefont {Li}}, \bibinfo {author} {\bibfnamefont {J.}~\bibnamefont
  {Liu}}, \bibinfo {author} {\bibfnamefont {J.~D.}\ \bibnamefont {McClain}},
  \bibinfo {author} {\bibfnamefont {E.~R.}\ \bibnamefont {Sayfutyarova}},
  \bibinfo {author} {\bibfnamefont {S.}~\bibnamefont {Sharma}}, \bibinfo
  {author} {\bibfnamefont {S.}~\bibnamefont {Wouters}}, \ and\ \bibinfo
  {author} {\bibfnamefont {G.~K.-L.}\ \bibnamefont {Chan}},\ }\href {\doibase
  10.1002/wcms.1340} {\bibfield  {journal} {\bibinfo  {journal} {Wiley
  Interdisciplinary Reviews: Computational Molecular Science}\ }\textbf
  {\bibinfo {volume} {8}},\ \bibinfo {pages} {e1340}}\BibitemShut {NoStop}%
\bibitem [{\citenamefont {Blankenbecler}\ \emph {et~al.}(1981)\citenamefont
  {Blankenbecler}, \citenamefont {Scalapino},\ and\ \citenamefont
  {Sugar}}]{Blankenbecler_PRD_1981}%
  \BibitemOpen
  \bibfield  {author} {\bibinfo {author} {\bibfnamefont {R.}~\bibnamefont
  {Blankenbecler}}, \bibinfo {author} {\bibfnamefont {D.~J.}\ \bibnamefont
  {Scalapino}}, \ and\ \bibinfo {author} {\bibfnamefont {R.~L.}\ \bibnamefont
  {Sugar}},\ }\href {\doibase 10.1103/PhysRevD.24.2278} {\bibfield  {journal}
  {\bibinfo  {journal} {Phys. Rev. D}\ }\textbf {\bibinfo {volume} {24}},\
  \bibinfo {pages} {2278} (\bibinfo {year} {1981})}\BibitemShut {NoStop}%
\bibitem [{\citenamefont {Sugiyama}\ and\ \citenamefont
  {Koonin}(1986)}]{Sugiyama_Annals_1986}%
  \BibitemOpen
  \bibfield  {author} {\bibinfo {author} {\bibfnamefont {G.}~\bibnamefont
  {Sugiyama}}\ and\ \bibinfo {author} {\bibfnamefont {S.}~\bibnamefont
  {Koonin}},\ }\href {\doibase https://doi.org/10.1016/0003-4916(86)90107-7}
  {\bibfield  {journal} {\bibinfo  {journal} {Annals of Physics}\ }\textbf
  {\bibinfo {volume} {168}},\ \bibinfo {pages} {1 } (\bibinfo {year}
  {1986})}\BibitemShut {NoStop}%
\bibitem [{\citenamefont {Zhang}\ \emph {et~al.}(1997)\citenamefont {Zhang},
  \citenamefont {Carlson},\ and\ \citenamefont
  {Gubernatis}}]{Zhang_PRB55_1997}%
  \BibitemOpen
  \bibfield  {author} {\bibinfo {author} {\bibfnamefont {S.}~\bibnamefont
  {Zhang}}, \bibinfo {author} {\bibfnamefont {J.}~\bibnamefont {Carlson}}, \
  and\ \bibinfo {author} {\bibfnamefont {J.~E.}\ \bibnamefont {Gubernatis}},\
  }\href {\doibase 10.1103/PhysRevB.55.7464} {\bibfield  {journal} {\bibinfo
  {journal} {Phys. Rev. B}\ }\textbf {\bibinfo {volume} {55}},\ \bibinfo
  {pages} {7464} (\bibinfo {year} {1997})}\BibitemShut {NoStop}%
\bibitem [{\citenamefont {Rom}\ \emph {et~al.}(1998)\citenamefont {Rom},
  \citenamefont {Fattal}, \citenamefont {Gupta}, \citenamefont {Carter},\ and\
  \citenamefont {Neuhauser}}]{Rom_JCP_1998}%
  \BibitemOpen
  \bibfield  {author} {\bibinfo {author} {\bibfnamefont {N.}~\bibnamefont
  {Rom}}, \bibinfo {author} {\bibfnamefont {E.}~\bibnamefont {Fattal}},
  \bibinfo {author} {\bibfnamefont {A.~K.}\ \bibnamefont {Gupta}}, \bibinfo
  {author} {\bibfnamefont {E.~A.}\ \bibnamefont {Carter}}, \ and\ \bibinfo
  {author} {\bibfnamefont {D.}~\bibnamefont {Neuhauser}},\ }\href {\doibase
  10.1063/1.477486} {\bibfield  {journal} {\bibinfo  {journal} {The Journal of
  Chemical Physics}\ }\textbf {\bibinfo {volume} {109}},\ \bibinfo {pages}
  {8241} (\bibinfo {year} {1998})}\BibitemShut {NoStop}%
\bibitem [{\citenamefont {Baer}\ \emph {et~al.}(1998)\citenamefont {Baer},
  \citenamefont {Head-Gordon},\ and\ \citenamefont
  {Neuhauser}}]{Baer_JCP_1998}%
  \BibitemOpen
  \bibfield  {author} {\bibinfo {author} {\bibfnamefont {R.}~\bibnamefont
  {Baer}}, \bibinfo {author} {\bibfnamefont {M.}~\bibnamefont {Head-Gordon}}, \
  and\ \bibinfo {author} {\bibfnamefont {D.}~\bibnamefont {Neuhauser}},\ }\href
  {\doibase 10.1063/1.477300} {\bibfield  {journal} {\bibinfo  {journal} {The
  Journal of Chemical Physics}\ }\textbf {\bibinfo {volume} {109}},\ \bibinfo
  {pages} {6219} (\bibinfo {year} {1998})}\BibitemShut {NoStop}%
\bibitem [{\citenamefont {Zhang}\ and\ \citenamefont
  {Krakauer}(2003)}]{Zhang_PRL90_2003}%
  \BibitemOpen
  \bibfield  {author} {\bibinfo {author} {\bibfnamefont {S.}~\bibnamefont
  {Zhang}}\ and\ \bibinfo {author} {\bibfnamefont {H.}~\bibnamefont
  {Krakauer}},\ }\href {\doibase 10.1103/PhysRevLett.90.136401} {\bibfield
  {journal} {\bibinfo  {journal} {Phys. Rev. Lett.}\ }\textbf {\bibinfo
  {volume} {90}},\ \bibinfo {pages} {136401} (\bibinfo {year}
  {2003})}\BibitemShut {NoStop}%
\bibitem [{\citenamefont {Al-Saidi}\ \emph {et~al.}(2006)\citenamefont
  {Al-Saidi}, \citenamefont {Zhang},\ and\ \citenamefont
  {Krakauer}}]{AlSaidi_JCP124_2006}%
  \BibitemOpen
  \bibfield  {author} {\bibinfo {author} {\bibfnamefont {W.~A.}\ \bibnamefont
  {Al-Saidi}}, \bibinfo {author} {\bibfnamefont {S.}~\bibnamefont {Zhang}}, \
  and\ \bibinfo {author} {\bibfnamefont {H.}~\bibnamefont {Krakauer}},\ }\href
  {\doibase 10.1063/1.2200885} {\bibfield  {journal} {\bibinfo  {journal} {J.
  Chem. Phys.}\ }\textbf {\bibinfo {volume} {124}},\ \bibinfo {pages} {224101}
  (\bibinfo {year} {2006})}\BibitemShut {NoStop}%
\bibitem [{\citenamefont {Zhang}(2013)}]{Zhang_Notes_2013}%
  \BibitemOpen
  \bibfield  {author} {\bibinfo {author} {\bibfnamefont {S.}~\bibnamefont
  {Zhang}},\ }in\ \href@noop {} {\emph {\bibinfo {booktitle} {Emergent
  Phenomena in Correlated Matter: Modeling and Simulation}}},\ \bibinfo
  {editor} {edited by\ \bibinfo {editor} {\bibfnamefont {E.~P.~E.}\
  \bibnamefont {Koch}}\ and\ \bibinfo {editor} {\bibfnamefont {U.}~\bibnamefont
  {Schollw\"ock}}}\ (\bibinfo  {publisher} {Verlag des Forschungszentrum
  J\"ulich},\ \bibinfo {year} {2013})\ Chap.~\bibinfo {chapter}
  {15}\BibitemShut {NoStop}%
\bibitem [{\citenamefont {Motta}\ and\ \citenamefont
  {Zhang}(2018{\natexlab{a}})}]{Motta_WIRES_2018}%
  \BibitemOpen
  \bibfield  {author} {\bibinfo {author} {\bibfnamefont {M.}~\bibnamefont
  {Motta}}\ and\ \bibinfo {author} {\bibfnamefont {S.}~\bibnamefont {Zhang}},\
  }\href {\doibase 10.1002/wcms.1364} {\bibfield  {journal} {\bibinfo
  {journal} {WIREs Comput Mol Sci}\ }\textbf {\bibinfo {volume} {e1364}},\
  \bibinfo {pages} {1} (\bibinfo {year} {2018}{\natexlab{a}})}\BibitemShut
  {NoStop}%
\bibitem [{\citenamefont {Evarestov}(2013)}]{Evarestov_book}%
  \BibitemOpen
  \bibfield  {author} {\bibinfo {author} {\bibfnamefont {R.}~\bibnamefont
  {Evarestov}},\ }\href@noop {} {\emph {\bibinfo {title} {Quantum Chemistry of
  Solids: LCAO Treatment of Crystals and Nanostructures}}},\ Springer Series in
  Solid-State Sciences\ (\bibinfo  {publisher} {Springer Berlin Heidelberg},\
  \bibinfo {year} {2013})\BibitemShut {NoStop}%
\bibitem [{\citenamefont {Dovesi}\ \emph {et~al.}(2014)\citenamefont {Dovesi},
  \citenamefont {Orlando}, \citenamefont {Erba}, \citenamefont
  {Zicovich-Wilson}, \citenamefont {Civalleri}, \citenamefont {Casassa},
  \citenamefont {Maschio}, \citenamefont {Ferrabone}, \citenamefont
  {De~La~Pierre}, \citenamefont {D'Arco}, \citenamefont {No\"{e}l},
  \citenamefont {Causa}, \citenamefont {Rerat},\ and\ \citenamefont
  {Kirtman}}]{dovesi2014crystal14}%
  \BibitemOpen
  \bibfield  {author} {\bibinfo {author} {\bibfnamefont {R.}~\bibnamefont
  {Dovesi}}, \bibinfo {author} {\bibfnamefont {R.}~\bibnamefont {Orlando}},
  \bibinfo {author} {\bibfnamefont {A.}~\bibnamefont {Erba}}, \bibinfo {author}
  {\bibfnamefont {C.~M.}\ \bibnamefont {Zicovich-Wilson}}, \bibinfo {author}
  {\bibfnamefont {B.}~\bibnamefont {Civalleri}}, \bibinfo {author}
  {\bibfnamefont {S.}~\bibnamefont {Casassa}}, \bibinfo {author} {\bibfnamefont
  {L.}~\bibnamefont {Maschio}}, \bibinfo {author} {\bibfnamefont
  {M.}~\bibnamefont {Ferrabone}}, \bibinfo {author} {\bibfnamefont
  {M.}~\bibnamefont {De~La~Pierre}}, \bibinfo {author} {\bibfnamefont
  {P.}~\bibnamefont {D'Arco}}, \bibinfo {author} {\bibfnamefont
  {Y.}~\bibnamefont {No\"{e}l}}, \bibinfo {author} {\bibfnamefont
  {M.}~\bibnamefont {Causa}}, \bibinfo {author} {\bibfnamefont
  {M.}~\bibnamefont {Rerat}}, \ and\ \bibinfo {author} {\bibfnamefont
  {B.}~\bibnamefont {Kirtman}},\ }\href {\doibase 10.1002/qua.24658} {\bibfield
   {journal} {\bibinfo  {journal} {International Journal of Quantum Chemistry}\
  }\textbf {\bibinfo {volume} {114}},\ \bibinfo {pages} {1287} (\bibinfo {year}
  {2014})}\BibitemShut {NoStop}%
\bibitem [{\citenamefont {Booth}\ \emph {et~al.}(2016)\citenamefont {Booth},
  \citenamefont {Tsatsoulis}, \citenamefont {Chan},\ and\ \citenamefont
  {Gr\"{u}neis}}]{booth2016plane}%
  \BibitemOpen
  \bibfield  {author} {\bibinfo {author} {\bibfnamefont {G.~H.}\ \bibnamefont
  {Booth}}, \bibinfo {author} {\bibfnamefont {T.}~\bibnamefont {Tsatsoulis}},
  \bibinfo {author} {\bibfnamefont {G.~K.-L.}\ \bibnamefont {Chan}}, \ and\
  \bibinfo {author} {\bibfnamefont {A.}~\bibnamefont {Gr\"{u}neis}},\ }\href
  {\doibase 10.1063/1.4961301} {\bibfield  {journal} {\bibinfo  {journal} {The
  Journal of Chemical Physics}\ }\textbf {\bibinfo {volume} {145}},\ \bibinfo
  {pages} {084111} (\bibinfo {year} {2016})}\BibitemShut {NoStop}%
\bibitem [{\citenamefont {Sun}\ \emph {et~al.}(2017)\citenamefont {Sun},
  \citenamefont {Berkelbach}, \citenamefont {McClain},\ and\ \citenamefont
  {Chan}}]{sun2017gaussian}%
  \BibitemOpen
  \bibfield  {author} {\bibinfo {author} {\bibfnamefont {Q.}~\bibnamefont
  {Sun}}, \bibinfo {author} {\bibfnamefont {T.~C.}\ \bibnamefont {Berkelbach}},
  \bibinfo {author} {\bibfnamefont {J.~D.}\ \bibnamefont {McClain}}, \ and\
  \bibinfo {author} {\bibfnamefont {G.~K.-L.}\ \bibnamefont {Chan}},\ }\href
  {\doibase 10.1063/1.4998644} {\bibfield  {journal} {\bibinfo  {journal} {The
  Journal of Chemical Physics}\ }\textbf {\bibinfo {volume} {147}},\ \bibinfo
  {pages} {164119} (\bibinfo {year} {2017})}\BibitemShut {NoStop}%
\bibitem [{\citenamefont {Gruber}\ \emph {et~al.}(2018)\citenamefont {Gruber},
  \citenamefont {Liao}, \citenamefont {Tsatsoulis}, \citenamefont {Hummel},\
  and\ \citenamefont {Gr\"uneis}}]{gruber2018applying}%
  \BibitemOpen
  \bibfield  {author} {\bibinfo {author} {\bibfnamefont {T.}~\bibnamefont
  {Gruber}}, \bibinfo {author} {\bibfnamefont {K.}~\bibnamefont {Liao}},
  \bibinfo {author} {\bibfnamefont {T.}~\bibnamefont {Tsatsoulis}}, \bibinfo
  {author} {\bibfnamefont {F.}~\bibnamefont {Hummel}}, \ and\ \bibinfo {author}
  {\bibfnamefont {A.}~\bibnamefont {Gr\"uneis}},\ }\href {\doibase
  10.1103/PhysRevX.8.021043} {\bibfield  {journal} {\bibinfo  {journal} {Phys.
  Rev. X}\ }\textbf {\bibinfo {volume} {8}},\ \bibinfo {pages} {021043}
  (\bibinfo {year} {2018})}\BibitemShut {NoStop}%
\bibitem [{\citenamefont {Lin}\ \emph {et~al.}(2001)\citenamefont {Lin},
  \citenamefont {Zong},\ and\ \citenamefont {Ceperley}}]{lin2001twist}%
  \BibitemOpen
  \bibfield  {author} {\bibinfo {author} {\bibfnamefont {C.}~\bibnamefont
  {Lin}}, \bibinfo {author} {\bibfnamefont {F.~H.}\ \bibnamefont {Zong}}, \
  and\ \bibinfo {author} {\bibfnamefont {D.~M.}\ \bibnamefont {Ceperley}},\
  }\href {\doibase 10.1103/PhysRevE.64.016702} {\bibfield  {journal} {\bibinfo
  {journal} {Phys. Rev. E}\ }\textbf {\bibinfo {volume} {64}},\ \bibinfo
  {pages} {016702} (\bibinfo {year} {2001})}\BibitemShut {NoStop}%
\bibitem [{\citenamefont {Ma}\ \emph {et~al.}(2015)\citenamefont {Ma},
  \citenamefont {Purwanto}, \citenamefont {Zhang},\ and\ \citenamefont
  {Krakauer}}]{Ma_PRL114_2015}%
  \BibitemOpen
  \bibfield  {author} {\bibinfo {author} {\bibfnamefont {F.}~\bibnamefont
  {Ma}}, \bibinfo {author} {\bibfnamefont {W.}~\bibnamefont {Purwanto}},
  \bibinfo {author} {\bibfnamefont {S.}~\bibnamefont {Zhang}}, \ and\ \bibinfo
  {author} {\bibfnamefont {H.}~\bibnamefont {Krakauer}},\ }\href {\doibase
  10.1103/PhysRevLett.114.226401} {\bibfield  {journal} {\bibinfo  {journal}
  {Phys. Rev. Lett.}\ }\textbf {\bibinfo {volume} {114}},\ \bibinfo {pages}
  {226401} (\bibinfo {year} {2015})}\BibitemShut {NoStop}%
\bibitem [{\citenamefont {Zhang}\ \emph {et~al.}(2018)\citenamefont {Zhang},
  \citenamefont {Malone},\ and\ \citenamefont {Morales}}]{Morales_arxiv_2018a}%
  \BibitemOpen
  \bibfield  {author} {\bibinfo {author} {\bibfnamefont {S.}~\bibnamefont
  {Zhang}}, \bibinfo {author} {\bibfnamefont {F.~D.}\ \bibnamefont {Malone}}, \
  and\ \bibinfo {author} {\bibfnamefont {M.~A.}\ \bibnamefont {Morales}},\
  }\href {\doibase 10.1063/1.5040900} {\bibfield  {journal} {\bibinfo
  {journal} {The Journal of Chemical Physics}\ }\textbf {\bibinfo {volume}
  {149}},\ \bibinfo {pages} {164102} (\bibinfo {year} {2018})}\BibitemShut
  {NoStop}%
\bibitem [{\citenamefont {Malone}\ \emph {et~al.}(2019)\citenamefont {Malone},
  \citenamefont {Zhang},\ and\ \citenamefont {Morales}}]{Morales_arxiv_2018b}%
  \BibitemOpen
  \bibfield  {author} {\bibinfo {author} {\bibfnamefont {F.~D.}\ \bibnamefont
  {Malone}}, \bibinfo {author} {\bibfnamefont {S.}~\bibnamefont {Zhang}}, \
  and\ \bibinfo {author} {\bibfnamefont {M.~A.}\ \bibnamefont {Morales}},\
  }\href {\doibase 10.1021/acs.jctc.8b00944} {\bibfield  {journal} {\bibinfo
  {journal} {Journal of Chemical Theory and Computation}\ }\textbf {\bibinfo
  {volume} {15}},\ \bibinfo {pages} {256} (\bibinfo {year} {2019})}\BibitemShut
  {NoStop}%
\bibitem [{\citenamefont {Hungerford}(1980)}]{Hungerford_book_1980}%
  \BibitemOpen
  \bibfield  {author} {\bibinfo {author} {\bibfnamefont {T.~W.}\ \bibnamefont
  {Hungerford}},\ }\href@noop {} {\emph {\bibinfo {title} {Algebra}}}\
  (\bibinfo  {publisher} {Springer},\ \bibinfo {year} {1980})\BibitemShut
  {NoStop}%
\bibitem [{\citenamefont {Dummit}\ and\ \citenamefont
  {Foote}(2004)}]{Dummit_book_2004}%
  \BibitemOpen
  \bibfield  {author} {\bibinfo {author} {\bibfnamefont {D.~S.}\ \bibnamefont
  {Dummit}}\ and\ \bibinfo {author} {\bibfnamefont {R.~M.}\ \bibnamefont
  {Foote}},\ }\href@noop {} {\emph {\bibinfo {title} {Abstract algebra}}}\
  (\bibinfo  {publisher} {New York: Wiley},\ \bibinfo {year}
  {2004})\BibitemShut {NoStop}%
\bibitem [{\citenamefont {Rudin}(1962)}]{Rudin_book_1962}%
  \BibitemOpen
  \bibfield  {author} {\bibinfo {author} {\bibfnamefont {W.}~\bibnamefont
  {Rudin}},\ }\href@noop {} {\emph {\bibinfo {title} {Fourier Analysis on
  Groups}}}\ (\bibinfo  {publisher} {Wiley},\ \bibinfo {year}
  {1962})\BibitemShut {NoStop}%
\bibitem [{\citenamefont {Jozsa}(1998)}]{Jozsa_RS_1998}%
  \BibitemOpen
  \bibfield  {author} {\bibinfo {author} {\bibfnamefont {R.}~\bibnamefont
  {Jozsa}},\ }\href {\doibase 10.1098/rspa.1998.0163} {\bibfield  {journal}
  {\bibinfo  {journal} {Proceedings of the Royal Society of London A:
  Mathematical, Physical and Engineering Sciences}\ }\textbf {\bibinfo {volume}
  {454}},\ \bibinfo {pages} {323} (\bibinfo {year} {1998})}\BibitemShut
  {NoStop}%
\bibitem [{\citenamefont {Purwanto}\ and\ \citenamefont
  {Zhang}(2004)}]{Purwanto_PRE70_2004}%
  \BibitemOpen
  \bibfield  {author} {\bibinfo {author} {\bibfnamefont {W.}~\bibnamefont
  {Purwanto}}\ and\ \bibinfo {author} {\bibfnamefont {S.}~\bibnamefont
  {Zhang}},\ }\href {\doibase 10.1103/PhysRevE.70.056702} {\bibfield  {journal}
  {\bibinfo  {journal} {Phys. Rev. E}\ }\textbf {\bibinfo {volume} {70}},\
  \bibinfo {pages} {056702} (\bibinfo {year} {2004})}\BibitemShut {NoStop}%
\bibitem [{\citenamefont {Hubbard}(1959)}]{Hubbard_PRL3_1959}%
  \BibitemOpen
  \bibfield  {author} {\bibinfo {author} {\bibfnamefont {J.}~\bibnamefont
  {Hubbard}},\ }\href {\doibase 10.1103/PhysRevLett.3.77} {\bibfield  {journal}
  {\bibinfo  {journal} {Phys. Rev. Lett.}\ }\textbf {\bibinfo {volume} {3}},\
  \bibinfo {pages} {77} (\bibinfo {year} {1959})}\BibitemShut {NoStop}%
\bibitem [{\citenamefont {Stratonovich}(1958)}]{Stratonovich_SPD2_1958}%
  \BibitemOpen
  \bibfield  {author} {\bibinfo {author} {\bibfnamefont {R.~L.}\ \bibnamefont
  {Stratonovich}},\ }\href@noop {} {\bibfield  {journal} {\bibinfo  {journal}
  {Soviet Physics Doklady}\ }\textbf {\bibinfo {volume} {2}},\ \bibinfo {pages}
  {416} (\bibinfo {year} {1958})}\BibitemShut {NoStop}%
\bibitem [{\citenamefont {LeBlanc}\ \emph {et~al.}(2015)\citenamefont
  {LeBlanc}, \citenamefont {Antipov}, \citenamefont {Becca}, \citenamefont
  {Bulik}, \citenamefont {Chan}, \citenamefont {Chung}, \citenamefont {Deng},
  \citenamefont {Ferrero}, \citenamefont {Henderson}, \citenamefont
  {Jim\'enez-Hoyos}, \citenamefont {Kozik}, \citenamefont {Liu}, \citenamefont
  {Millis}, \citenamefont {Prokof'ev}, \citenamefont {Qin}, \citenamefont
  {Scuseria}, \citenamefont {Shi}, \citenamefont {Svistunov}, \citenamefont
  {Tocchio}, \citenamefont {Tupitsyn}, \citenamefont {White}, \citenamefont
  {Zhang}, \citenamefont {Zheng}, \citenamefont {Zhu},\ and\ \citenamefont
  {Gull}}]{LeBlanc_PRX5_2015}%
  \BibitemOpen
  \bibfield  {author} {\bibinfo {author} {\bibfnamefont {J.~P.~F.}\
  \bibnamefont {LeBlanc}}, \bibinfo {author} {\bibfnamefont {A.~E.}\
  \bibnamefont {Antipov}}, \bibinfo {author} {\bibfnamefont {F.}~\bibnamefont
  {Becca}}, \bibinfo {author} {\bibfnamefont {I.~W.}\ \bibnamefont {Bulik}},
  \bibinfo {author} {\bibfnamefont {G.~K.-L.}\ \bibnamefont {Chan}}, \bibinfo
  {author} {\bibfnamefont {C.-M.}\ \bibnamefont {Chung}}, \bibinfo {author}
  {\bibfnamefont {Y.}~\bibnamefont {Deng}}, \bibinfo {author} {\bibfnamefont
  {M.}~\bibnamefont {Ferrero}}, \bibinfo {author} {\bibfnamefont {T.~M.}\
  \bibnamefont {Henderson}}, \bibinfo {author} {\bibfnamefont {C.~A.}\
  \bibnamefont {Jim\'enez-Hoyos}}, \bibinfo {author} {\bibfnamefont
  {E.}~\bibnamefont {Kozik}}, \bibinfo {author} {\bibfnamefont {X.-W.}\
  \bibnamefont {Liu}}, \bibinfo {author} {\bibfnamefont {A.~J.}\ \bibnamefont
  {Millis}}, \bibinfo {author} {\bibfnamefont {N.~V.}\ \bibnamefont
  {Prokof'ev}}, \bibinfo {author} {\bibfnamefont {M.}~\bibnamefont {Qin}},
  \bibinfo {author} {\bibfnamefont {G.~E.}\ \bibnamefont {Scuseria}}, \bibinfo
  {author} {\bibfnamefont {H.}~\bibnamefont {Shi}}, \bibinfo {author}
  {\bibfnamefont {B.~V.}\ \bibnamefont {Svistunov}}, \bibinfo {author}
  {\bibfnamefont {L.~F.}\ \bibnamefont {Tocchio}}, \bibinfo {author}
  {\bibfnamefont {I.~S.}\ \bibnamefont {Tupitsyn}}, \bibinfo {author}
  {\bibfnamefont {S.~R.}\ \bibnamefont {White}}, \bibinfo {author}
  {\bibfnamefont {S.}~\bibnamefont {Zhang}}, \bibinfo {author} {\bibfnamefont
  {B.-X.}\ \bibnamefont {Zheng}}, \bibinfo {author} {\bibfnamefont
  {Z.}~\bibnamefont {Zhu}}, \ and\ \bibinfo {author} {\bibfnamefont
  {E.}~\bibnamefont {Gull}} (\bibinfo {collaboration} {Simons Collaboration on
  the Many-Electron Problem}),\ }\href {\doibase 10.1103/PhysRevX.5.041041}
  {\bibfield  {journal} {\bibinfo  {journal} {Phys. Rev. X}\ }\textbf {\bibinfo
  {volume} {5}},\ \bibinfo {pages} {041041} (\bibinfo {year}
  {2015})}\BibitemShut {NoStop}%
\bibitem [{\citenamefont {Qin}\ \emph {et~al.}(2016)\citenamefont {Qin},
  \citenamefont {Shi},\ and\ \citenamefont {Zhang}}]{Qin_PRB_2016}%
  \BibitemOpen
  \bibfield  {author} {\bibinfo {author} {\bibfnamefont {M.}~\bibnamefont
  {Qin}}, \bibinfo {author} {\bibfnamefont {H.}~\bibnamefont {Shi}}, \ and\
  \bibinfo {author} {\bibfnamefont {S.}~\bibnamefont {Zhang}},\ }\href
  {\doibase 10.1103/PhysRevB.94.235119} {\bibfield  {journal} {\bibinfo
  {journal} {Phys. Rev. B}\ }\textbf {\bibinfo {volume} {94}},\ \bibinfo
  {pages} {235119} (\bibinfo {year} {2016})}\BibitemShut {NoStop}%
\bibitem [{\citenamefont {Zheng}\ \emph {et~al.}(2017)\citenamefont {Zheng},
  \citenamefont {Chung}, \citenamefont {Corboz}, \citenamefont {Ehlers},
  \citenamefont {Qin}, \citenamefont {Noack}, \citenamefont {Shi},
  \citenamefont {White}, \citenamefont {Zhang},\ and\ \citenamefont
  {Chan}}]{Zheng_Science_2017}%
  \BibitemOpen
  \bibfield  {author} {\bibinfo {author} {\bibfnamefont {B.-X.}\ \bibnamefont
  {Zheng}}, \bibinfo {author} {\bibfnamefont {C.-M.}\ \bibnamefont {Chung}},
  \bibinfo {author} {\bibfnamefont {P.}~\bibnamefont {Corboz}}, \bibinfo
  {author} {\bibfnamefont {G.}~\bibnamefont {Ehlers}}, \bibinfo {author}
  {\bibfnamefont {M.-P.}\ \bibnamefont {Qin}}, \bibinfo {author} {\bibfnamefont
  {R.~M.}\ \bibnamefont {Noack}}, \bibinfo {author} {\bibfnamefont
  {H.}~\bibnamefont {Shi}}, \bibinfo {author} {\bibfnamefont {S.~R.}\
  \bibnamefont {White}}, \bibinfo {author} {\bibfnamefont {S.}~\bibnamefont
  {Zhang}}, \ and\ \bibinfo {author} {\bibfnamefont {G.~K.-L.}\ \bibnamefont
  {Chan}},\ }\href {\doibase 10.1126/science.aam7127} {\bibfield  {journal}
  {\bibinfo  {journal} {Science}\ }\textbf {\bibinfo {volume} {358}},\ \bibinfo
  {pages} {1155} (\bibinfo {year} {2017})}\BibitemShut {NoStop}%
\bibitem [{\citenamefont {Purwanto}\ \emph {et~al.}(2015)\citenamefont
  {Purwanto}, \citenamefont {Zhang},\ and\ \citenamefont
  {Krakauer}}]{Purwanto2014}%
  \BibitemOpen
  \bibfield  {author} {\bibinfo {author} {\bibfnamefont {W.}~\bibnamefont
  {Purwanto}}, \bibinfo {author} {\bibfnamefont {S.}~\bibnamefont {Zhang}}, \
  and\ \bibinfo {author} {\bibfnamefont {H.}~\bibnamefont {Krakauer}},\ }\href
  {\doibase 10.1063/1.4906829} {\bibfield  {journal} {\bibinfo  {journal} {The
  Journal of Chemical Physics}\ }\textbf {\bibinfo {volume} {142}},\ \bibinfo
  {pages} {064302} (\bibinfo {year} {2015})}\BibitemShut {NoStop}%
\bibitem [{\citenamefont {Motta}\ \emph {et~al.}(2017)\citenamefont {Motta},
  \citenamefont {Ceperley}, \citenamefont {Chan}, \citenamefont {Gomez},
  \citenamefont {Gull}, \citenamefont {Guo}, \citenamefont {Jim\'enez-Hoyos},
  \citenamefont {Lan}, \citenamefont {Li}, \citenamefont {Ma}, \citenamefont
  {Millis}, \citenamefont {Prokof'ev}, \citenamefont {Ray}, \citenamefont
  {Scuseria}, \citenamefont {Sorella}, \citenamefont {Stoudenmire},
  \citenamefont {Sun}, \citenamefont {Tupitsyn}, \citenamefont {White},
  \citenamefont {Zgid},\ and\ \citenamefont {Zhang}}]{Motta_PRX_2017}%
  \BibitemOpen
  \bibfield  {author} {\bibinfo {author} {\bibfnamefont {M.}~\bibnamefont
  {Motta}}, \bibinfo {author} {\bibfnamefont {D.~M.}\ \bibnamefont {Ceperley}},
  \bibinfo {author} {\bibfnamefont {G.~K.-L.}\ \bibnamefont {Chan}}, \bibinfo
  {author} {\bibfnamefont {J.~A.}\ \bibnamefont {Gomez}}, \bibinfo {author}
  {\bibfnamefont {E.}~\bibnamefont {Gull}}, \bibinfo {author} {\bibfnamefont
  {S.}~\bibnamefont {Guo}}, \bibinfo {author} {\bibfnamefont {C.~A.}\
  \bibnamefont {Jim\'enez-Hoyos}}, \bibinfo {author} {\bibfnamefont {T.~N.}\
  \bibnamefont {Lan}}, \bibinfo {author} {\bibfnamefont {J.}~\bibnamefont
  {Li}}, \bibinfo {author} {\bibfnamefont {F.}~\bibnamefont {Ma}}, \bibinfo
  {author} {\bibfnamefont {A.~J.}\ \bibnamefont {Millis}}, \bibinfo {author}
  {\bibfnamefont {N.~V.}\ \bibnamefont {Prokof'ev}}, \bibinfo {author}
  {\bibfnamefont {U.}~\bibnamefont {Ray}}, \bibinfo {author} {\bibfnamefont
  {G.~E.}\ \bibnamefont {Scuseria}}, \bibinfo {author} {\bibfnamefont
  {S.}~\bibnamefont {Sorella}}, \bibinfo {author} {\bibfnamefont {E.~M.}\
  \bibnamefont {Stoudenmire}}, \bibinfo {author} {\bibfnamefont
  {Q.}~\bibnamefont {Sun}}, \bibinfo {author} {\bibfnamefont {I.~S.}\
  \bibnamefont {Tupitsyn}}, \bibinfo {author} {\bibfnamefont {S.~R.}\
  \bibnamefont {White}}, \bibinfo {author} {\bibfnamefont {D.}~\bibnamefont
  {Zgid}}, \ and\ \bibinfo {author} {\bibfnamefont {S.}~\bibnamefont {Zhang}}
  (\bibinfo {collaboration} {Simons Collaboration on the Many-Electron
  Problem}),\ }\href {\doibase 10.1103/PhysRevX.7.031059} {\bibfield  {journal}
  {\bibinfo  {journal} {Phys. Rev. X}\ }\textbf {\bibinfo {volume} {7}},\
  \bibinfo {pages} {031059} (\bibinfo {year} {2017})}\BibitemShut {NoStop}%
\bibitem [{\citenamefont {Shee}\ \emph {et~al.}(2019)\citenamefont {Shee},
  \citenamefont {Rudshteyn}, \citenamefont {Arthur}, \citenamefont {Zhang},
  \citenamefont {Reichman},\ and\ \citenamefont {Friesner}}]{Shee_JCTC_2019}%
  \BibitemOpen
  \bibfield  {author} {\bibinfo {author} {\bibfnamefont {J.}~\bibnamefont
  {Shee}}, \bibinfo {author} {\bibfnamefont {B.}~\bibnamefont {Rudshteyn}},
  \bibinfo {author} {\bibfnamefont {E.~J.}\ \bibnamefont {Arthur}}, \bibinfo
  {author} {\bibfnamefont {S.}~\bibnamefont {Zhang}}, \bibinfo {author}
  {\bibfnamefont {D.~R.}\ \bibnamefont {Reichman}}, \ and\ \bibinfo {author}
  {\bibfnamefont {R.~A.}\ \bibnamefont {Friesner}},\ }\href {\doibase
  10.1021/acs.jctc.9b00083} {\bibfield  {journal} {\bibinfo  {journal} {Journal
  of Chemical Theory and Computation}\ }\textbf {\bibinfo {volume} {15}},\
  \bibinfo {pages} {2346} (\bibinfo {year} {2019})},\ \bibinfo {note} {pMID:
  30883110}\BibitemShut {NoStop}%
\bibitem [{\citenamefont {Motta}\ and\ \citenamefont
  {Zhang}(2017)}]{Motta_JCTC_2017}%
  \BibitemOpen
  \bibfield  {author} {\bibinfo {author} {\bibfnamefont {M.}~\bibnamefont
  {Motta}}\ and\ \bibinfo {author} {\bibfnamefont {S.}~\bibnamefont {Zhang}},\
  }\href {\doibase 10.1021/acs.jctc.7b00730} {\bibfield  {journal} {\bibinfo
  {journal} {Journal of Chemical Theory and Computation}\ }\textbf {\bibinfo
  {volume} {13}},\ \bibinfo {pages} {5367} (\bibinfo {year} {2017})},\ \bibinfo
  {note} {pMID: 29053270}\BibitemShut {NoStop}%
\bibitem [{\citenamefont {Motta}\ and\ \citenamefont
  {Zhang}(2018{\natexlab{b}})}]{Motta_JCP_2017}%
  \BibitemOpen
  \bibfield  {author} {\bibinfo {author} {\bibfnamefont {M.}~\bibnamefont
  {Motta}}\ and\ \bibinfo {author} {\bibfnamefont {S.}~\bibnamefont {Zhang}},\
  }\href {\doibase 10.1063/1.5029508} {\bibfield  {journal} {\bibinfo
  {journal} {The Journal of Chemical Physics}\ }\textbf {\bibinfo {volume}
  {148}},\ \bibinfo {pages} {181101} (\bibinfo {year}
  {2018}{\natexlab{b}})}\BibitemShut {NoStop}%
\bibitem [{\citenamefont {Shee}\ \emph {et~al.}(2017)\citenamefont {Shee},
  \citenamefont {Zhang}, \citenamefont {Reichman},\ and\ \citenamefont
  {Friesner}}]{Shee_JCTC13_2017}%
  \BibitemOpen
  \bibfield  {author} {\bibinfo {author} {\bibfnamefont {J.}~\bibnamefont
  {Shee}}, \bibinfo {author} {\bibfnamefont {S.}~\bibnamefont {Zhang}},
  \bibinfo {author} {\bibfnamefont {D.~R.}\ \bibnamefont {Reichman}}, \ and\
  \bibinfo {author} {\bibfnamefont {R.~A.}\ \bibnamefont {Friesner}},\ }\href
  {\doibase 10.1021/acs.jctc.7b00224} {\bibfield  {journal} {\bibinfo
  {journal} {J. Chem. Theor. Comput.}\ }\textbf {\bibinfo {volume} {13}},\
  \bibinfo {pages} {2667} (\bibinfo {year} {2017})}\BibitemShut {NoStop}%
\bibitem [{\citenamefont {Motta}\ \emph {et~al.}(2018)\citenamefont {Motta},
  \citenamefont {Shee}, \citenamefont {Zhang},\ and\ \citenamefont
  {Chan}}]{Motta_arxiv_2019}%
  \BibitemOpen
  \bibfield  {author} {\bibinfo {author} {\bibfnamefont {M.}~\bibnamefont
  {Motta}}, \bibinfo {author} {\bibfnamefont {J.}~\bibnamefont {Shee}},
  \bibinfo {author} {\bibfnamefont {S.}~\bibnamefont {Zhang}}, \ and\ \bibinfo
  {author} {\bibfnamefont {G.~K.-L.}\ \bibnamefont {Chan}},\ }\href@noop {}
  (\bibinfo {year} {2018}),\ \Eprint
  {http://arxiv.org/abs/arXiv:1810.01549} {arXiv:1810.01549} \BibitemShut
  {NoStop}%
\bibitem [{\citenamefont {Whitten}(1973)}]{DF1}%
  \BibitemOpen
  \bibfield  {author} {\bibinfo {author} {\bibfnamefont {J.~L.}\ \bibnamefont
  {Whitten}},\ }\href {\doibase 10.1063/1.1679012} {\bibfield  {journal}
  {\bibinfo  {journal} {The Journal of Chemical Physics}\ }\textbf {\bibinfo
  {volume} {58}},\ \bibinfo {pages} {4496} (\bibinfo {year}
  {1973})}\BibitemShut {NoStop}%
\bibitem [{\citenamefont {Hohenstein}\ and\ \citenamefont
  {Sherrill}(2010)}]{DF2}%
  \BibitemOpen
  \bibfield  {author} {\bibinfo {author} {\bibfnamefont {E.~G.}\ \bibnamefont
  {Hohenstein}}\ and\ \bibinfo {author} {\bibfnamefont {C.~D.}\ \bibnamefont
  {Sherrill}},\ }\href {\doibase 10.1063/1.3426316} {\bibfield  {journal}
  {\bibinfo  {journal} {The Journal of Chemical Physics}\ }\textbf {\bibinfo
  {volume} {132}},\ \bibinfo {pages} {184111} (\bibinfo {year}
  {2010})}\BibitemShut {NoStop}%
\bibitem [{\citenamefont {Beebe}\ and\ \citenamefont
  {Linderberg}(1977)}]{Beebe_IJQC12_1977}%
  \BibitemOpen
  \bibfield  {author} {\bibinfo {author} {\bibfnamefont {N.~H.~F.}\
  \bibnamefont {Beebe}}\ and\ \bibinfo {author} {\bibfnamefont
  {J.}~\bibnamefont {Linderberg}},\ }\href {\doibase 10.1002/qua.560120408}
  {\bibfield  {journal} {\bibinfo  {journal} {International Journal of Quantum
  Chemistry}\ }\textbf {\bibinfo {volume} {12}},\ \bibinfo {pages} {683}
  (\bibinfo {year} {1977})}\BibitemShut {NoStop}%
\bibitem [{\citenamefont {Koch}\ \emph {et~al.}(2003)\citenamefont {Koch},
  \citenamefont {de~Meras},\ and\ \citenamefont {Pedersen}}]{Koch_JCP118_2003}%
  \BibitemOpen
  \bibfield  {author} {\bibinfo {author} {\bibfnamefont {H.}~\bibnamefont
  {Koch}}, \bibinfo {author} {\bibfnamefont {A.~S.}\ \bibnamefont {de~Meras}},
  \ and\ \bibinfo {author} {\bibfnamefont {T.~B.}\ \bibnamefont {Pedersen}},\
  }\href {\doibase 10.1063/1.1578621} {\bibfield  {journal} {\bibinfo
  {journal} {J. Chem. Phys.}\ }\textbf {\bibinfo {volume} {118}},\ \bibinfo
  {pages} {9481} (\bibinfo {year} {2003})}\BibitemShut {NoStop}%
\bibitem [{\citenamefont {Aquilante}\ \emph {et~al.}(2010)\citenamefont
  {Aquilante}, \citenamefont {De~Vico}, \citenamefont {Ferr\'e}, \citenamefont
  {Ghigo}, \citenamefont {Malmqvist}, \citenamefont {Neogr\'ady}, \citenamefont
  {Pedersen}, \citenamefont {Piton\'ak}, \citenamefont {Reiher}, \citenamefont
  {Roos}, \citenamefont {Serrano-Andr\'es}, \citenamefont {Urban},
  \citenamefont {Veryazov},\ and\ \citenamefont
  {Lindh}}]{Aquilante_JCC31_2010}%
  \BibitemOpen
  \bibfield  {author} {\bibinfo {author} {\bibfnamefont {F.}~\bibnamefont
  {Aquilante}}, \bibinfo {author} {\bibfnamefont {L.}~\bibnamefont {De~Vico}},
  \bibinfo {author} {\bibfnamefont {N.}~\bibnamefont {Ferr\'e}}, \bibinfo
  {author} {\bibfnamefont {G.}~\bibnamefont {Ghigo}}, \bibinfo {author}
  {\bibfnamefont {P.-A.}\ \bibnamefont {Malmqvist}}, \bibinfo {author}
  {\bibfnamefont {P.}~\bibnamefont {Neogr\'ady}}, \bibinfo {author}
  {\bibfnamefont {T.~B.}\ \bibnamefont {Pedersen}}, \bibinfo {author}
  {\bibfnamefont {M.}~\bibnamefont {Piton\'ak}}, \bibinfo {author}
  {\bibfnamefont {M.}~\bibnamefont {Reiher}}, \bibinfo {author} {\bibfnamefont
  {B.~O.}\ \bibnamefont {Roos}}, \bibinfo {author} {\bibfnamefont
  {L.}~\bibnamefont {Serrano-Andr\'es}}, \bibinfo {author} {\bibfnamefont
  {M.}~\bibnamefont {Urban}}, \bibinfo {author} {\bibfnamefont
  {V.}~\bibnamefont {Veryazov}}, \ and\ \bibinfo {author} {\bibfnamefont
  {R.}~\bibnamefont {Lindh}},\ }\href {\doibase 10.1002/jcc.21318} {\bibfield
  {journal} {\bibinfo  {journal} {Journal of Computational Chemistry}\ }\textbf
  {\bibinfo {volume} {31}},\ \bibinfo {pages} {224} (\bibinfo {year}
  {2010})}\BibitemShut {NoStop}%
\bibitem [{\citenamefont {Purwanto}\ \emph {et~al.}(2011)\citenamefont
  {Purwanto}, \citenamefont {Krakauer}, \citenamefont {Virgus},\ and\
  \citenamefont {Zhang}}]{Purwanto_JCP135_2011}%
  \BibitemOpen
  \bibfield  {author} {\bibinfo {author} {\bibfnamefont {W.}~\bibnamefont
  {Purwanto}}, \bibinfo {author} {\bibfnamefont {H.}~\bibnamefont {Krakauer}},
  \bibinfo {author} {\bibfnamefont {Y.}~\bibnamefont {Virgus}}, \ and\ \bibinfo
  {author} {\bibfnamefont {S.}~\bibnamefont {Zhang}},\ }\href {\doibase
  10.1063/1.3654002} {\bibfield  {journal} {\bibinfo  {journal} {J. Chem.
  Phys.}\ }\textbf {\bibinfo {volume} {135}},\ \bibinfo {pages} {164105}
  (\bibinfo {year} {2011})}\BibitemShut {NoStop}%
\bibitem [{\citenamefont {Wick}(1950)}]{Wick_PR80_1950}%
  \BibitemOpen
  \bibfield  {author} {\bibinfo {author} {\bibfnamefont {G.~C.}\ \bibnamefont
  {Wick}},\ }\href {\doibase 10.1103/PhysRev.80.268} {\bibfield  {journal}
  {\bibinfo  {journal} {Phys. Rev.}\ }\textbf {\bibinfo {volume} {80}},\
  \bibinfo {pages} {268} (\bibinfo {year} {1950})}\BibitemShut {NoStop}%
\bibitem [{\citenamefont {Balian}\ and\ \citenamefont
  {Brezin}(1969)}]{Balian_NC64_1969}%
  \BibitemOpen
  \bibfield  {author} {\bibinfo {author} {\bibfnamefont {R.}~\bibnamefont
  {Balian}}\ and\ \bibinfo {author} {\bibfnamefont {E.}~\bibnamefont
  {Brezin}},\ }\href {\doibase 10.1007/BF02710281} {\bibfield  {journal}
  {\bibinfo  {journal} {Nuovo Cimento B}\ }\textbf {\bibinfo {volume} {64}},\
  \bibinfo {pages} {37} (\bibinfo {year} {1969})}\BibitemShut {NoStop}%
\bibitem [{\citenamefont {Dunning}(1989)}]{dunning1989cc1}%
  \BibitemOpen
  \bibfield  {author} {\bibinfo {author} {\bibfnamefont {T.~H.}\ \bibnamefont
  {Dunning}},\ }\href {\doibase 10.1063/1.456153} {\bibfield  {journal}
  {\bibinfo  {journal} {The Journal of Chemical Physics}\ }\textbf {\bibinfo
  {volume} {90}},\ \bibinfo {pages} {1007} (\bibinfo {year}
  {1989})}\BibitemShut {NoStop}%
\bibitem [{\citenamefont {Woon}\ and\ \citenamefont
  {Dunning}(1993)}]{dunning1989cc3}%
  \BibitemOpen
  \bibfield  {author} {\bibinfo {author} {\bibfnamefont {D.~E.}\ \bibnamefont
  {Woon}}\ and\ \bibinfo {author} {\bibfnamefont {T.~H.}\ \bibnamefont
  {Dunning}},\ }\href {\doibase 10.1063/1.464303} {\bibfield  {journal}
  {\bibinfo  {journal} {The Journal of Chemical Physics}\ }\textbf {\bibinfo
  {volume} {98}},\ \bibinfo {pages} {1358} (\bibinfo {year}
  {1993})}\BibitemShut {NoStop}%
\bibitem [{\citenamefont {Goedecker}\ \emph {et~al.}(1996)\citenamefont
  {Goedecker}, \citenamefont {Teter},\ and\ \citenamefont
  {Hutter}}]{Goedecker_PRB_1996}%
  \BibitemOpen
  \bibfield  {author} {\bibinfo {author} {\bibfnamefont {S.}~\bibnamefont
  {Goedecker}}, \bibinfo {author} {\bibfnamefont {M.}~\bibnamefont {Teter}}, \
  and\ \bibinfo {author} {\bibfnamefont {J.}~\bibnamefont {Hutter}},\ }\href
  {\doibase 10.1103/PhysRevB.54.1703} {\bibfield  {journal} {\bibinfo
  {journal} {Phys. Rev. B}\ }\textbf {\bibinfo {volume} {54}},\ \bibinfo
  {pages} {1703} (\bibinfo {year} {1996})}\BibitemShut {NoStop}%
\bibitem [{\citenamefont {Hartwigsen}\ \emph {et~al.}(1998)\citenamefont
  {Hartwigsen}, \citenamefont {Goedecker},\ and\ \citenamefont
  {Hutter}}]{Hartwigsen_PRB_1998}%
  \BibitemOpen
  \bibfield  {author} {\bibinfo {author} {\bibfnamefont {C.}~\bibnamefont
  {Hartwigsen}}, \bibinfo {author} {\bibfnamefont {S.}~\bibnamefont
  {Goedecker}}, \ and\ \bibinfo {author} {\bibfnamefont {J.}~\bibnamefont
  {Hutter}},\ }\href {\doibase 10.1103/PhysRevB.58.3641} {\bibfield  {journal}
  {\bibinfo  {journal} {Phys. Rev. B}\ }\textbf {\bibinfo {volume} {58}},\
  \bibinfo {pages} {3641} (\bibinfo {year} {1998})}\BibitemShut {NoStop}%
\bibitem [{\citenamefont {Hutter}\ \emph {et~al.}(2014)\citenamefont {Hutter},
  \citenamefont {Iannuzzi}, \citenamefont {Schiffmann},\ and\ \citenamefont
  {VandeVondele}}]{cp2kbasisref}%
  \BibitemOpen
  \bibfield  {author} {\bibinfo {author} {\bibfnamefont {J.}~\bibnamefont
  {Hutter}}, \bibinfo {author} {\bibfnamefont {M.}~\bibnamefont {Iannuzzi}},
  \bibinfo {author} {\bibfnamefont {F.}~\bibnamefont {Schiffmann}}, \ and\
  \bibinfo {author} {\bibfnamefont {J.}~\bibnamefont {VandeVondele}},\ }\href
  {\doibase 10.1002/wcms.1159} {\bibfield  {journal} {\bibinfo  {journal}
  {Wiley Interdisciplinary Reviews: Computational Molecular Science}\ }\textbf
  {\bibinfo {volume} {4}},\ \bibinfo {pages} {15} (\bibinfo {year}
  {2014})}\BibitemShut {NoStop}%
\bibitem [{\citenamefont {Kwee}\ \emph {et~al.}(2008)\citenamefont {Kwee},
  \citenamefont {Zhang},\ and\ \citenamefont {Krakauer}}]{Kwee_PRL100_2008}%
  \BibitemOpen
  \bibfield  {author} {\bibinfo {author} {\bibfnamefont {H.}~\bibnamefont
  {Kwee}}, \bibinfo {author} {\bibfnamefont {S.}~\bibnamefont {Zhang}}, \ and\
  \bibinfo {author} {\bibfnamefont {H.}~\bibnamefont {Krakauer}},\ }\href
  {\doibase 10.1103/PhysRevLett.100.126404} {\bibfield  {journal} {\bibinfo
  {journal} {Phys. Rev. Lett.}\ }\textbf {\bibinfo {volume} {100}},\ \bibinfo
  {pages} {126404} (\bibinfo {year} {2008})}\BibitemShut {NoStop}%
\bibitem [{\citenamefont {Schimka}\ \emph {et~al.}(2011)\citenamefont
  {Schimka}, \citenamefont {Harl},\ and\ \citenamefont
  {Kresse}}]{Schima_JCP_2011}%
  \BibitemOpen
  \bibfield  {author} {\bibinfo {author} {\bibfnamefont {L.}~\bibnamefont
  {Schimka}}, \bibinfo {author} {\bibfnamefont {J.}~\bibnamefont {Harl}}, \
  and\ \bibinfo {author} {\bibfnamefont {G.}~\bibnamefont {Kresse}},\ }\href
  {\doibase 10.1063/1.3524336} {\bibfield  {journal} {\bibinfo  {journal} {The
  Journal of Chemical Physics}\ }\textbf {\bibinfo {volume} {134}},\ \bibinfo
  {pages} {024116} (\bibinfo {year} {2011})}\BibitemShut {NoStop}%
\bibitem [{Note1()}]{Note1}%
  \BibitemOpen
  \bibinfo {note} {\protect \tmspace +\thinmuskip
  {.1667em}https://github.com/cryos/avogadro/blob/master/\\crystals/nitrides/BN.cif}\BibitemShut
  {NoStop}%
\end{thebibliography}

%

\end{document}